\documentclass[twocolumn,amsmath,amssymb,aps]{revtex4-2}
\usepackage{lineno,hyperref}
\usepackage{bm}
\usepackage{amsmath}
\usepackage{gensymb}
\usepackage{epstopdf}
\usepackage{booktabs}
\usepackage{float}

\usepackage{graphicx}
\usepackage{ulem}
\usepackage{xcolor}
\begin{document}

\title{4\textit{d}-element induced improvement of structural disorder and development of weakly re-entrant spin-glass behaviour in NiRuMnSn}

\author{Shuvankar Gupta$^{1,6}$}
\author{Sudip Chakraborty$^{1,6}$}
\author{Vidha Bhasin$^{2,6}$}
\author{Santanu Pakhira$^3$}
\author{Anis Biswas$^3$}
\author{Yaroslav Mudryk$^3$}
\author{Amit Kumar$^{5,6}$}
\author{Celine Barreteau$^4$}
\author{Jean-Claude Crivello$^4$}
\author{Amitabh Das$^{5,6}$}
\author{S.N. Jha$^2$}
\author{D. Bhattacharyya$^2$}
\author{Vitalij K. Pecharsky$^{3,7}$}
\thanks{deceased}
\author{Eric Alleno$^4$}
\author{Chandan Mazumdar$^{1}$}
\email{chandan.mazumdar@saha.in}

\affiliation{$^1$Condensed Matter Physics Division, Saha Institute of Nuclear Physics, 1/AF, Bidhannagar, Kolkata 700064, India}
\affiliation{$^2$Atomic \& Molecular Physics Division, Bhabha Atomic Research Centre, Mumbai 400 094, Maharashtra,  India}
\affiliation{$^3$Ames National Laboratory, Iowa State University, Ames, Iowa 50011, USA}
\affiliation{$^4$Universit\'{e} Paris-Est, Institut de Chimie et des Mat\'{e}riaux Paris-Est, UMR 7182 CNRS UPEC, 2 rue H. Dunant, 94320 Thiais, France}
\affiliation{$^5$Solid State Physics Division, Bhabha Atomic Research Centre, Mumbai 400 085, India}
\affiliation{$^6$Homi Bhabha National Institute, Training School Complex, Anushaktinagar, Mumbai 400094, India}
\affiliation{$^7$Department of Materials Science and Engineering, Iowa State University, Ames, Iowa 50011, USA}
\date{\today}
\begin{abstract}

The pursuit of efficient spin-polarization in quaternary Heusler alloys with the general formula $XX'YZ$ (where \textit{X}, $X'$, and \textit{Y} are transition metals and \textit{Z} is a \textit{p}-block element), has been a subject of significant scientific interest. While previous studies shows that isoelectronic substitution of 4\textit{d} element in place of 3\textit{d} element in quaternary Heusler alloy, improves the half-metallic ferromagnetic characteristics, our research on the quaternary Heusler alloy NiRuMnSn reveals a strikingly different scenario. In this study, we present a detailed structural analysis of the material using X-ray absorption fine structure (EXAFS) and neutron diffraction (ND) techniques, which confirms the formation of a single-phase compound with 50:50 site disorder between Ni/Ru atoms at 4\textit{c}/4\textit{d} sites. Contrary to expectations, our DFT calculations suggests a considerable decrease in spin-polarization even in the ordered structure. Additionally, we report on the compound's exceptional behavior, displaying a rare re-entrant spin glass property below $\sim$60\, K, a unique and intriguing feature for quaternary Heusler-type compounds.

\end{abstract}
\maketitle

\section{\label{sec:Introduction}Introduction}

An extended family of the so-called Heusler alloys continues to attract considerable attention of the condensed matter physics and materials science communities due to a plethora of tailorable properties that both are fundamentally interesting and potentially functional. Well-known examples include half-metallic ferromagnetism (HMF)~\cite{de1983new}, ferromagnetic shape memory effects~\cite{manosa2010giant}, formation of magnetic skyrmions ~\cite{nayak2017magnetic}, topological phenomena and Weyl semimetallicity~\cite{manna2018heusler}, unusual thermoelectricity~\cite{hinterleitner2019thermoelectric,mondal2018ferromagnetically}, giant magnetocaloric effect~\cite{liu2012giant}, and others. Generally, Heusler phases are classified as either full Heusler, commonly represented by the idealized $X_2YZ$ stoichiometries, where \textit{X} and \textit{Y} are transition elements and \textit{Z} is main-group element, or half Heulser, often quoted as XYZ compounds~\cite{graf2011simple}. Structurally ordered full Heusler alloys crystallize with the $L2_1$-type structure (space group: $Fm\bar{3}m$, no. 225) that consists of four interpenetrating face-centered cubic (fcc) lattices. For the case of half Heusler alloy, one of the fcc lattice sites remains vacant and it crystalizes in Y-type of crystal structure (Space group: $F\bar{4}3m$, no. 216)~\cite{graf2011simple}. A rather simple cubic crystal structure makes them an ideal model system for basic research focused on the fundamental understanding of \textit{d}-band magnetism~\cite{telling2008evidence,barth2010itinerant,roy2019complex}. Apart from this, half-metallic properties of many Heusler alloys can be useful in the field of spintronics~\cite{felser2007spintronics}, making use of high spin-polarization due to unique band structures of HFM Heusler alloys, in which one spin channel is metallic, whereas the other spin channel is semi-conducting in nature~\cite{bombor2013half}.

It is generally found that the degree of spin polarization in HMF systems is very sensitive to structural disorder~\cite{miura2004atomic,kharel2017effect,mukadam2016quantification,bera2022selective}. As materials that belong to the Heusler family often include elements from the same period of the periodic table (\textit{e.g.}, \textit{3d} elements) they crystallize in disordered structures, making it challenging to synthesize Heulser phases with negligible site-disorder. In order to tune the Fermi energy ($E_F$) in the middle of the gap, which makes HMF property more robust against external perturbation, partial substitutions at the \textit{Y}-site in full Heusler systems are often found to be effective. However, such substitution often results in considerable structural disorder, which, in turn, reduces spin polarization~\cite{bainsla2016equiatomic}. On the other hand, replacing half of the \textit{X} atoms by another transition element, {$X'$}, is possible, leading to quaternary Heulser alloys ($XX'YZ$)~\cite{dai2009new}. The latter crystalize in the \textit{Y}-type (LiMgPdSn-type) structure (space group: $F\bar{4}3m$, no. 216), in which each of the four interpenetrating lattices is occupied by four different elements~\cite{bainsla2016equiatomic}, which is achieved by splitting the 8\textit{c} site in $Fm\bar{3}m$ into the 4\textit{c} and 4\textit{d} sites in $F\bar{4}3m$. Research on quaternary Heusler alloy systems mainly focuses on 3\textit{d}-based transition elements~\cite{alijani2011electronic,alijani2011quaternary,bainsla2015spin,bainsla2015origin,gupta2022coexisting,gupta2023high,gupta2023rare}, whereas only a handful of 4\textit{d}-based (Ru, Rh) Heusler alloys are reported in the literature~\cite{rani2017structural,bainsla2015corufex}. 4\textit{d}-based Heusler alloys are known to form with negligibly small structural disorder and thus they show enhanced magnetic ordering temperatures without altering the basic nature of magnetic characteristics of their 3\textit{d}-based counterparts~\cite{bainsla2015corufex}. For example,
CoFeMnGe crystallizes in the $D0_3$-type disordered structure (3\textit{d}-elements are statistically distributed among 3 different sites) due to presence of elements of nearly equal size~\cite{bainsla2014high}. However, when Fe is replaced by Rh, the compound CoRhMnGe crystallizes in the \textit{Y}-type ordered structure~\cite{rani2017structural}.

NiFeMnSn is a 3\textit{d}-based quarternary Heusler phase that exhibits a moderately high Curie temperature, $T_{\rm C}$ $\sim$ 405\,K, which is theoretically claimed to possess a moderately high spin polarization of $\sim70$\% in the ordered structure~\cite{mukadam2016quantification,bera2022selective}. However, due to the presence of inherent structural disorder in the experimentally synthesized compound, the half-metallicity is considerably diminished, and the compound becomes metallic. An isoelectronic substitution of a 4\textit{d}-element Ru in place of 3\textit{d}-element Fe is expected to enhance the spin-polarization through improved structural ordering. Through our detailed experimental and theoretical investigation, we however find that despite achieving a much improved atomic periodicity in NiRuMnSn, spin-polarisation actually deteriorates, which we have assigned to the much weaker magnetic character of Ru when compared to Fe. This nonmagnetic (or weakly magnetic) character of Ru also results in the appearance of weakly reentrant spin-glass behaviour, which is very rare in equiatomic quaternary Heusler alloy family.

\section{Methods}
\subsection{Experimental}
The polycrystalline NiRuMnSn was prepared by arc melting of a stoichiometric mixture of high purity ($>$99.99\%) constituent elements in a flowing argon environment. To compensate for the losses due to the vaporization of Mn, the element was weighed with 2 wt.\% excess. To improve homogeneity, the alloy was remelted five times, flipping the button after each melting. The X-ray diffraction (XRD) analysis of powdered sample at room temperature was performed with Cu-K$\alpha$ radiation using a rotating anode TTRAX-III diffractometer (Rigaku Corp., Japan). The sample’s phase purity and crystal structure were determined by performing Rietveld refinement using the FullProf software package~\cite{rodriguez1993recent}. dc magnetization measurements were carried out in a commercial superconducting quantum interference device (SQUID)- magnetometer (Quantum Design Inc, USA) at temperatures 5$-$380 K under applied magnetic fields of  $-70 \leq H \leq 70$\,kOe. ac magnetic susceptibility measurements were performed in an MPMS XL-7 SQUID magnetometer (Quantum Design Inc., USA). Electrical resistivity was measured with a four-probe technique using a Physical Property Measurement System (Quantum Design Inc., USA). The extended X-ray absorption fine structure (EXAFS) measurements were performed at the BL-9 beamline at the Indus-2 Synchrotron Source (2.5 GeV, 100 mA) at Raja Ramanna Centre for Advanced Technology (RRCAT), Indore, India. Powder Neutron diffraction data were recorded on the PD2 powder neutron diffractometer ($\lambda$ = 1.2443 {\AA}) at the Dhruva reactor, Bhabha Atomic Research Centre (BARC), Mumbai, India. Neutron depolarization measurements were performed on the polarized neutron spectrometer at the Dhruva reactor.

\subsection{Computational}
The structural stability and ground-state properties of the system are theoretically investigated using the density functional theory (DFT). DFT calculations were conducted using the projector augmented wave (PAW) method~\cite{blochl1994projector} implemented in the Vienna ab initio simulation package (VASP)~\cite{kresse1993ab,kresse1994norm}. The exchange correlation was described by the generalized gradient approximation modified by Perdew, Burke and Ernzerhof (GGA-PBE)~\cite{perdew1996generalized}. Energy bands up to a cutoff energy, $E_{cutoff} = 600$\,eV, were included in all calculations. Upon performing volume and ionic (for disordered compounds) relaxation steps, the tetrahedron
method with Bl\"{o}chl correction~\cite{blochl1994improved} was applied. For all the cases, the spin-polarization calculations were appropriately considered. In order to model statistical chemical disorder in NiRuMnSn, unit cells based on the concept of special quasirandom structure (SQS)~\cite{zunger1990special} were generated to model different possible disorder schemes. To generate the SQS, the cluster expansion formalism for the multicomponent and multisublattice systems~\cite{sanchez1984generalized} was used as implemented in the Monte-Carlo (MCSQS) code contained in the Alloy-Theoretic Automated Toolkit (ATAT)~\cite{van2009multicomponent,van2013efficient}. Subsequent DFT calculations were performed in order to test the quality of the SQS and to see how reliable are the DFT results. The root mean square (\textit{rms}) error was used as another quality criterion besides the calculations including a different order of interactions. The \textit{rms} error describes the deviation of the correlation function of the SQS ($\Pi^{k}_{SQS}$) from the correlation functions of considered clusters k for a given random structure ($\Pi^{k}_{md}$)
\begin{equation}
\ rms = \sqrt{\sum_{k}(\Pi^{k}_{SQS}-\Pi^{k}_{md})^2}
\label{eq1}
\end{equation}

Several tests on the dependence of the cluster type and nearest neighbors numbers were done to generate the disordered structure with Sn in 4\textit{a} (0,0,0), Mn in 4\textit{b} (0.5,0.5,0.5), 0.5 Ru + 0.5 Ni in 4\textit{c} (0.25,0.25,0.25), and 0.5 Ru + 0.5 Ni in 4\textit{d} (0.75,0.75,0.75). Finally, seven first-pair, five triplet, and 11 quadruplet interactions were considered to obtain reliable results. For the 4\textit{c}/4\textit{d}-disordered \textit{Y}-phase, a quaternary SQS cell of 28 atoms was build.

\section{Results and Discussion}

\subsection{\label{sec:Electronic_Structure_Ordered} Structure optimization and electronic structure calculations}
\begin{figure}[ht]
\centerline{\includegraphics[width=.48\textwidth]{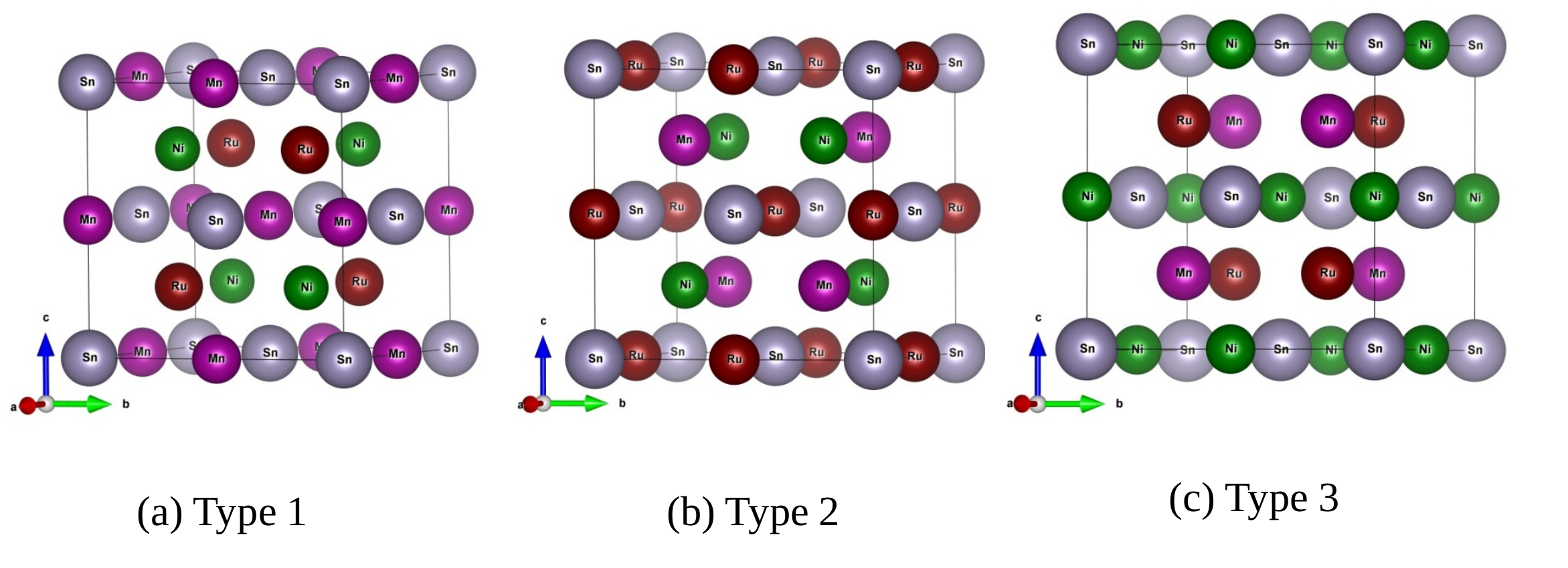}}
{\caption{Unit cell representation of (a) Type-1 (b) Type-2 and (c) Type-3 ordered structure as described in Table~\ref{Enthalpy}.
Sn, Mn, Ni and Ru atoms are represented by grey, pink, green and red balls respectivly.}\label{Fig_1}}
\end{figure}

DFT calculations on NiRuMnSn using different atomic arrangements within the LiMgPdSn-type structure were first performed to find the most stable configuration. In a quaternary Heusler alloy $XX'YZ$, assuming \textit{Z}-element atoms occupy the 4\textit{a} (0,0,0) site, the remaining three elements $X$, $X'$ and $Y$ could be placed on 4\textit{b} (0.5, 0.5, 0.5), 4\textit{c} (0.25, 0.25, 0.25) and 4\textit{d} (0.75, 0.75, 0.75) sites. Further, considering that switching atoms between the 4\textit{c} and 4\textit{d} sites results in energetically degenerate configurations, out of 6 total possible ordered NiRuMnSn structures, only three that are listed in Table~\ref{Enthalpy} and depicted in Fig.~\ref{Fig_1} are distinguishable.

\begin{table}
\caption{Calculated enthalpy of formation $\Delta_f{H}$ for different 3 ordered types and one disordered structural atomic arrangement of NiRuMnSn.}
\begin{tabular}{|c|c|c|c|c|c|}
\hline
 & 4\textit{a} & 4\textit{b} & 4\textit{c} & 4\textit{d} & $\Delta_f{H}$ $(kJ/mol)$ \\
\hline
Type-1 & Sn & Mn & Ni & Ru & -8.19\\
\hline
Type-2 & Sn & Ru & Mn & Ni & 14.80 \\
\hline
Type-3 & Sn & Ni & Ru & Mn & 8.95\\
\hline
disordered & Sn & Mn & Ni:Ru & Ni:Ru & -11.14 \\
\hline
\end{tabular}\\
\label{Enthalpy}
\end{table}

\begin{figure*}[]
\centerline{\includegraphics[width=.96\textwidth]{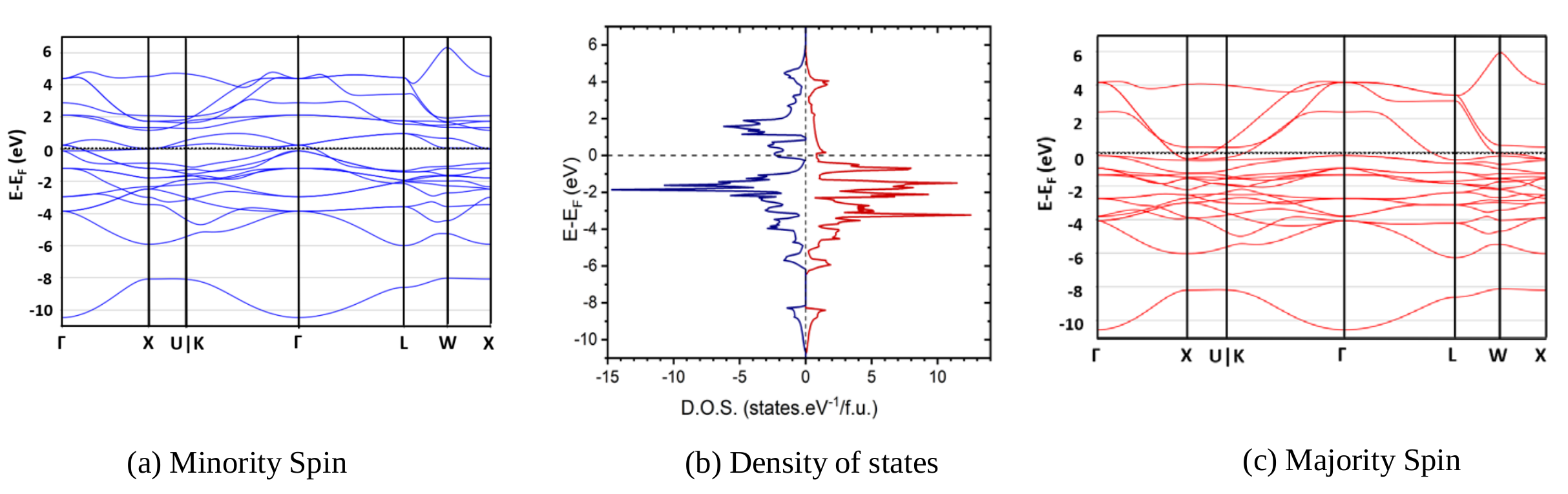}}
{\caption{Spin-polarized band structure and density of states of NiRuMnSn in ordered Type-1 structure: (a) minority spin band, (b) density of states and (c) majority spin band.}\label{Fig_2}}
\end{figure*}

The DFT-calculated enthalpies of formation presented in Table~\ref{Enthalpy} show that the ordered Type-1 structure (with Ni and Ru in the same planes) is more stable than Type-2 and Type-3, which is consistent with site preferences in other quaternary Heusler compounds, where the least electronegative atoms occupy the 4\textit{b} site. Figure~\ref{Fig_2} shows the calculated spin-polarized band structure and the density of states (DOS) of the energetically most favorable Type-1 configuration. We note that DOS reveals no band gaps in either for spin-up or for spin-down states. Additionally, the calculated spin polarization (44.6\%) is moderate, and the calculated total magnetic moment, 4.3 $\mu_B$ (Ni = 0.46 $\mu_B$, Ru = 0.31 $\mu_B$ and Mn = 3.53 $\mu_B$), is close to 5 $\mu_B$ predicted by the Slater-Pauling rule~\cite{galanakis2002slater,ozdougan2013slater}.

\subsection{\label{sec:XRD}X-ray diffraction}

\begin{figure}[h]
\centerline{\includegraphics[width=.48\textwidth]{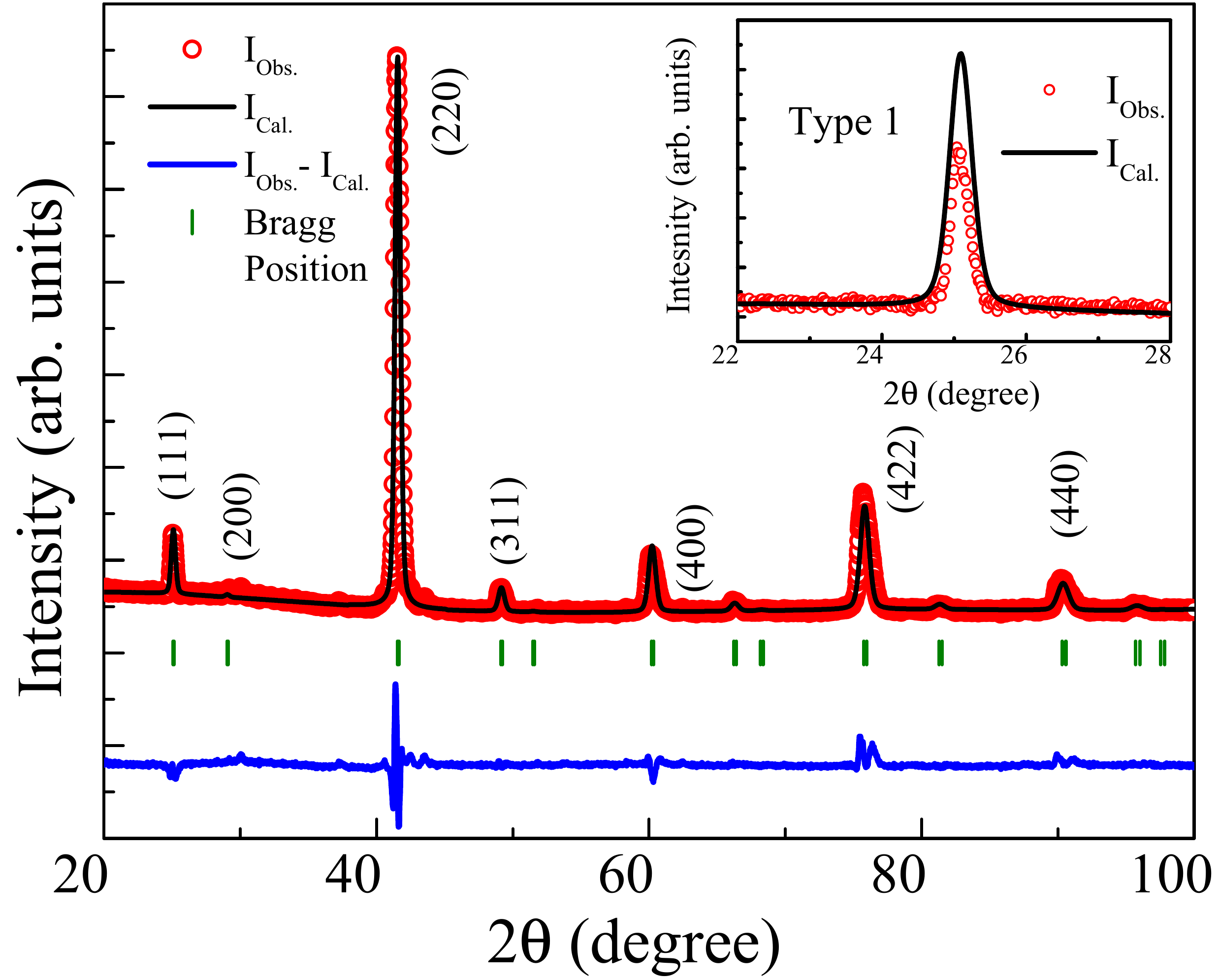}}
{\caption{Rietveld refinement of the powder XRD pattern of NiRuMnSn (assuming disordered structure) measured at room temperature. Inset highlights the Rietveld refinement results using the (111) peak as an example when considering the ordered structure (Type-1).}\label{Fig_3}}
\end{figure}

\begin{figure}[h]
\centerline{\includegraphics[width=.48\textwidth]{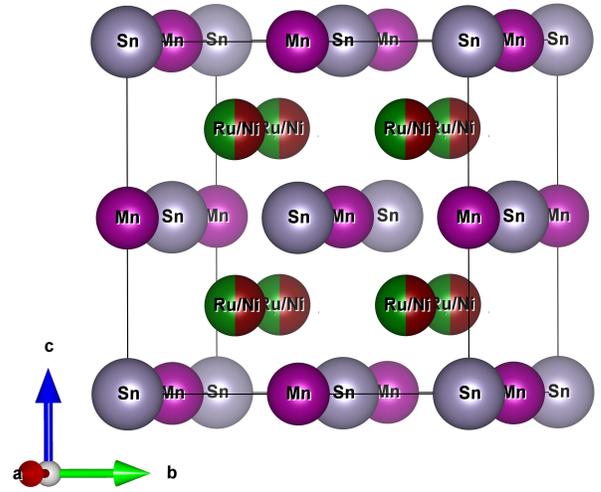}}
{\caption{Atomic arrangement in the disordered crystal structure of NiRuMnSn.}\label{Fig_4}}
\end{figure}

Theory predicts (Table~\ref{Enthalpy}) that Type-1 structure, in which Sn occupies the 4\textit{a}, Mn$-$4\textit{b}, Ru$-$4\textit{c}, and Ni$-$4\textit{d} positions, has lower energy compared to the other two types of ordered structures. In other words, the most stable ordered structure should consist of Sn/Mn and Ni/Ru layers. However, the Rietveld refinement of powder X-ray diffraction (XRD) pattern taken at room temperature does not support the formation of the fully-ordered Type-1 structure as the intensity of the (111) peak, as illustrated in the inset of Fig.~\ref{Fig_3}, could not be described satisfactorily using the atomic positions from Table~\ref{Enthalpy}. Since intensities of both the (111) and (200) Bragg peaks are most indicative of the presence of structural disorder, the mismatch between observed and calculated intensities of the (111) peak suggests presence of structural disorder in the Type-1 arrangement of crystal lattice~\cite{bainsla2016equiatomic,webster1973magnetic,gupta2022coexisting}. Out of different types of disorder present in Heusler alloys, \textit{A2} (space group: $Im\bar{3}m$, no. 229) and \textit{B2} (space group: $Pm\bar{3}m$, no. 221) types are known to occur most frequently. For the A2-type structure, both the (111) and (200) peaks are absent, whereas only the (200) peak is present in the \textit{B2}-type~\cite{graf2011simple,bainsla2016equiatomic,gupta2022coexisting}. In the \textit{A2}-type structure, all the constituent atoms ($X$, $X'$, $Y$ and $Z$) randomly mix with each other while for the \textit{B2}-type, the \textit{Y} \& \textit{Z} and \textit{X} \& $X'$ atoms randomly mix with each other in 4\textit{b} \& 4\textit{a} and 4\textit{c} \& 4\textit{d} sites, respectively. Clear presence of the (111) and near absence of the (200) Bragg peaks, therefore, contradicts both the \textit{A2}- and \textit{ B2}-type disorder in the title material.

A model in which Ni and Ru atoms mix with each other in 4\textit{c} and 4\textit{d} positions, on the other hand, leads to a satisfactory fit of intensities of all Bragg peaks present in the powder diffraction pattern (Fig.~\ref{Fig_3}). This disorder between the Ni and Ru atoms transforms the \textit{Y}-type ordered structure (space group: $F\bar{4}3m$, no. 216) into a variant of the $L2_1$-type (space group: $Fm\bar{3}m$) structure typical of ordered ternary Heusler compounds. The disordered crystal structure is presented in Fig.~\ref{Fig_4}. Subsequently, we have also checked the ground state energy for this disordered structure and found it has lower formation enthalpy (-11.14 kJ/mol ) compared to the ordered Type-1 structure (see Table.~\ref{Enthalpy}) consistent with experimental results. Similar type of disorder Co and Ru atoms was previously reported in CoRuMnSi~\cite{venkateswara2020half}. The lattice parameter determined by Rietveld refinement is ${a=b=c=}$ 6.145(4)\,{\AA}.\\

\subsection{\label{sec:EXAFS}Extended X-ray absorption fine structure (EXAFS)}
Details of crystal structure may be difficult to establish conclusively using X-ray powder diffraction alone, particularly when two elements with similar atomic scattering factors, in this case Mn and Ni, are present in the same lattice. Other experiments, such as, extended X-ray absorption fine structure (EXAFS) and neutron diffraction (ND)~\cite{balke2007structural,bainsla2015local,mukadam2016quantification,ravel2002exafs} are therefore, highly beneficial to determine the exact atomic arrangements in the title system. In contrast to XRD, EXAFS is an element-specific measurement that focuses on the local atomic environment around the absorbing atoms. Therefore, to better understand the nature of disorder, we have performed the EXAFS measurements at the Ni-, Ru- and Mn \textit{K}-edges on the BL-9 beam-line operating in the energy range 4–25 keV~\cite{basu2014comprehensive, poswal2014commissioning}. The collimating meridional cylindrical mirror is covered with Rh/Pt, and the collimated beam reflected by the mirror is monochromatized by a Si(111) (2\textit{d}=6.2709 {\AA}) based double crystal monochromator (DCM). The second DCM  is a sagittal cylinder that is utilised for horizontal focusing, and a Rh/Pt coated bendable post mirror facing down is used for vertical focusing at the sample position. Detuning the second crystal of DCM is used to reject the higher harmonics content in the X-ray beam. EXAFS measurements in both transmission and fluorescence mode are possible in this beamline. For this case, the measurements were done in the fluorescence mode, with the sample at 45 degrees to the incident X-ray beam and a fluorescence detector at right angles to the incident X-ray beam to capture the signal. The incident flux (I$_0$) is measured by one ionisation chamber detector, and the fluorescence intensity is measured by a florescence detector (I$_f$). The spectra were acquired as a function of energy by scanning the monochromator across the prescribed range, and the absorption coefficient ($\mu$) was obtained using the relation:  I$_T$ = I$_0$e$^{-{\mu}x}$, where x is the thickness of the absorber.

\begin{figure}
\begin{minipage}{0.49\textwidth}
\centering
{\includegraphics[width=.98\textwidth]{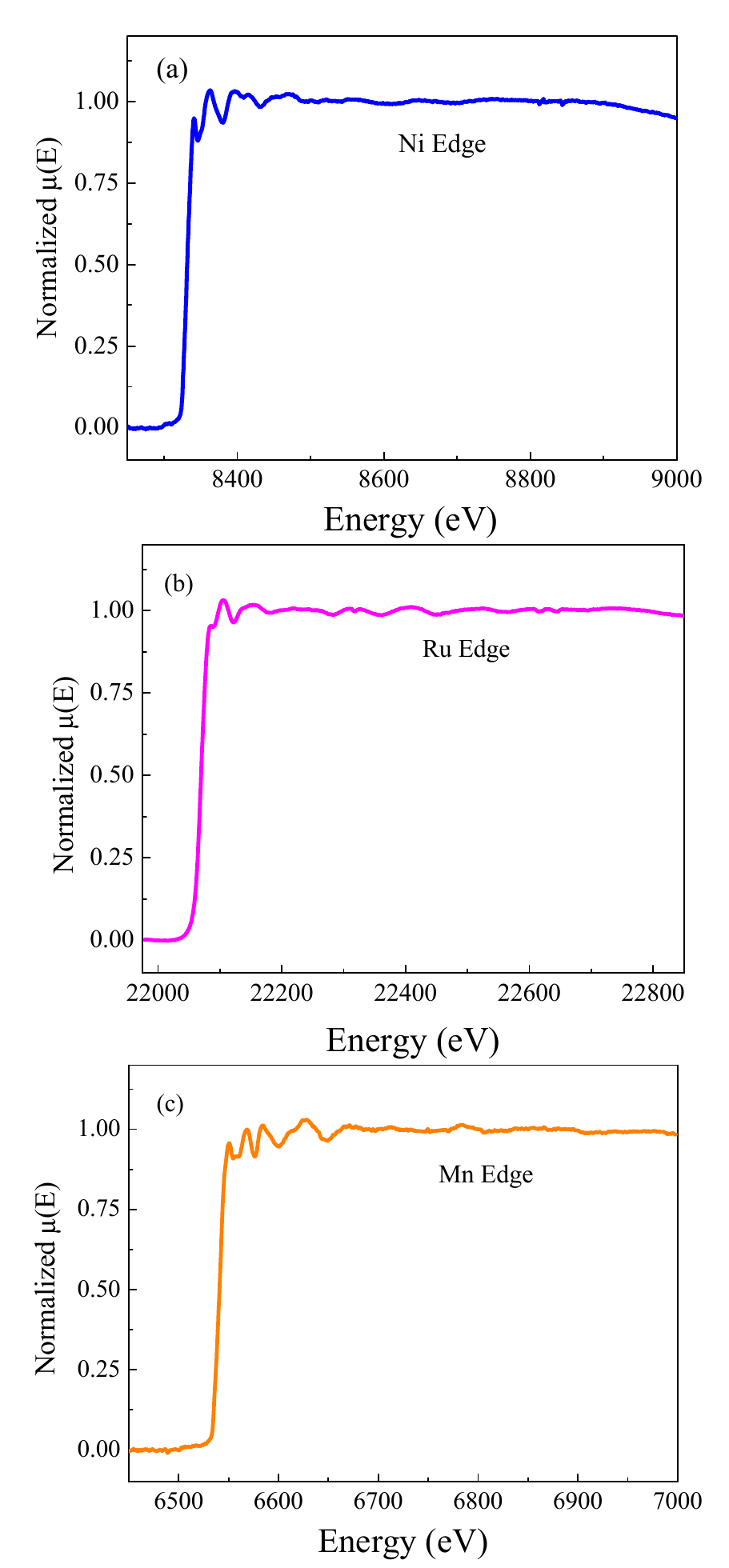}}
{\caption{Normalized EXAFS spectra of NiRuMnSn taken at  (a) Ni-edge, (b) Ru-edge, and (c) Mn-edge.}\label{Fig_5}}
\end{minipage}
\end{figure}

\begin{figure}
\begin{minipage}{0.49\textwidth}
\centering
{\includegraphics[width=.98\textwidth]{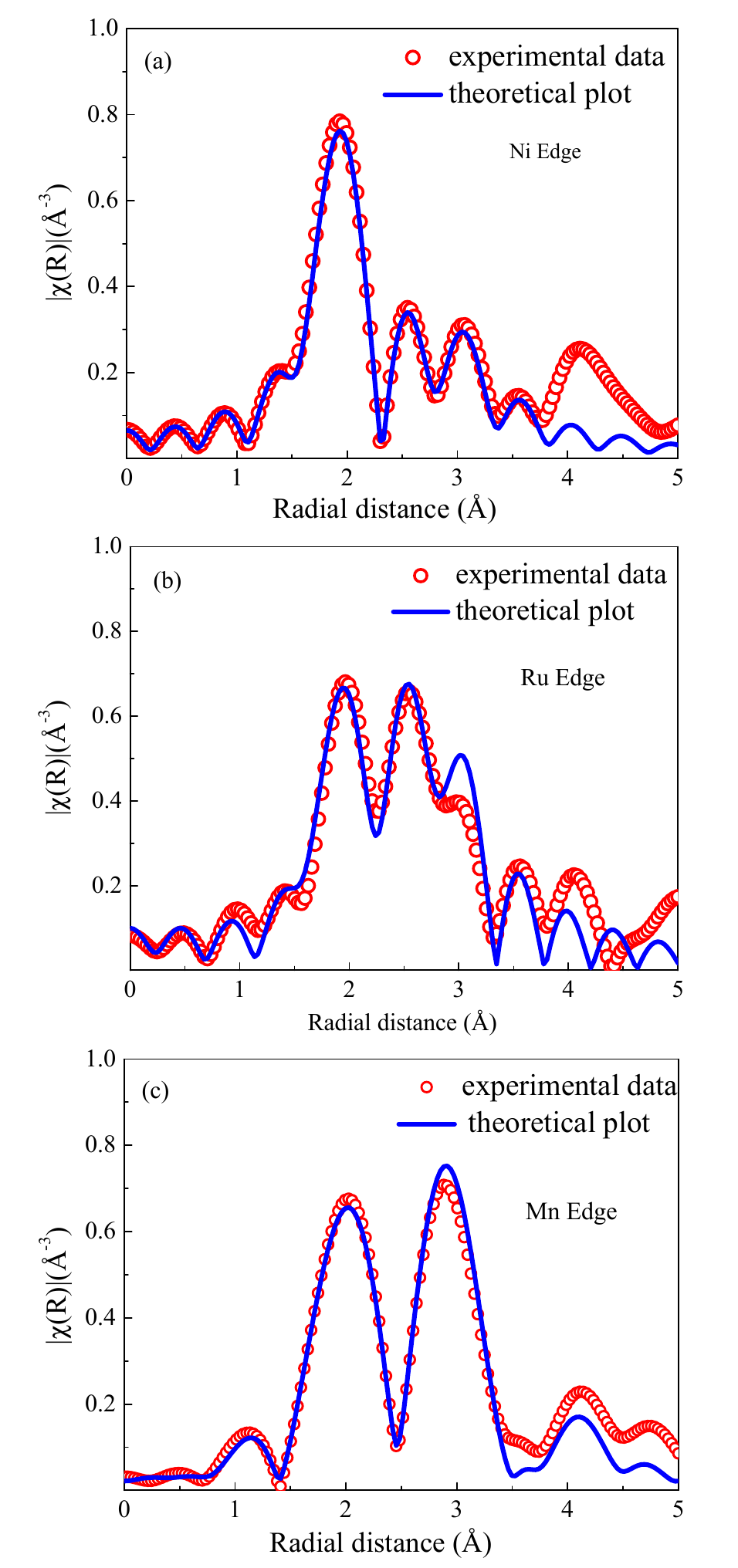}}
{\caption{Fourier transformed EXAFS spectra of NiRuMnSn taken at  (a) Ni-edge, (b) Ru-edge, and (c) Mn-edge (scatter red points) and theoretical fit (solid blue line).}\label{Fig_6}}
\end{minipage}
\end{figure}

Fig.~\ref{Fig_5} shows the normalised EXAFS ($\mu$(E) \textit{vs.} E) spectra of NiRuMnSn at the Ni-, Ru- and Mn- \textit{K}-edges. Oscillations in the absorption spectra ($\mu$(E) \textit{vs.} E) has been translated to EXAFS function $\chi$(E) defined as~\cite{konigsberger1988x}
\begin{equation}
\chi(E) = \frac{\mu(E)-\mu_{0}(E)} {\Delta\mu_{0}(E_{0})}
\label{eq1}
\end{equation}
\noindent
where E$_0$ is the absorption edge energy, $\mu$$_0$(E$_0$) is the bare atom background and $\Delta$$\mu$$_{0}$(E$_{0}$) is the absorption edge step in $\mu$(E) value. $\chi$(E) has been transformed to the wave number dependent absorption coefficient $\chi$(k) using the relation,
\begin{equation}
k=\sqrt {\frac {2m(E-E_{0})}{\hbar^{2}}}
\label{eq2}
\end{equation}
\noindent
where, \textit{m} is the electron mass, $\chi$(\textit{k}) is weighted by \textit{k$^2$} to amplify the oscillation at high \textit{k}  and the $\chi$(\textit{k})\textit{k}$^2$ functions are Fourier converted in \textit{R} space to obtain the $\chi$(R) \textit{vs.} R graphs in terms of real distances from the centre of the absorbing atom. For EXAFS data analysis, a suite of EXAFS data analysis programme included in the Demeter software package~\cite{ravel2005athena} were employed. This includes background reduction and Fourier transform to derive the $\chi$(\textit{R}) \textit{vs.} \textit{R} plots from absorption spectra (using ATHENA code), generation of theoretical EXAFS spectra based on an assumed crystallographic structure, and finally fitting experimental data to theoretical spectra using ARTEMIS code.

Fig.~\ref{Fig_6} shows the Fourier transformed EXAFS spectra ($\chi$(\textit{R}) \textit{vs.} \textit{R} ) plots of Ni-, Ru- and Mn-edges for NiRuMnSn. Simultaneous fitting of various data sets with numerous edges is the fitting approach used here ~\cite{ravel2002exafs}. The $\chi$(\textit{R}) \textit{vs.} \textit{R} plots measured at Ni-, Ru- and Mn- edges are fitted simultaneously with common fitting parameters. The number of independent parameters is reduced below the Nyquist requirements, and the statistical significance of the fitted model is improved. The goodness of the fit determined by the value of the R$_{factor}$ is described by
\begin{equation}
 \scalebox{1} {$R_{factor}= \frac{[Im(\chi_{dat}(r_{i})-\chi_{th}(r_{i})]^{2} + [Re((\chi_{dat}(r_{i})-\chi_{th}(r_{i})]^{2}]^{2}} {[Im(\chi_{dat}(r_i)^{2}]+[Re(\chi_{dat}(r_i)^{2}]}$}
\label{eq3}
\end{equation}
\noindent
where ${\chi_{dat}}$ and ${\chi_{th}}$ are the experimental and theoretical ${\chi(R)}$ values, \textit{Im} and \textit{Re} are the imaginary and real components of the related quantities, respectively. For theoretical simulation of the EXAFS spectra of NiRuMnSn, the structural parameters have been obtained from the XRD results. Bond distances (\textit{R}), co-ordination numbers (N), and disorder (Debye-Waller) factors (${\sigma^2}$), which give the mean square variations in the distances, were employed as fitting parameters during the fitting process. The fitting has been done up to 2.8 {\AA} for Ni- and Ru- edge data and up to 3.8 {\AA} for Mn-edge data.

\begin{table*}[ht]
\caption{Bond length (R), coordination number (N), and Debye-Waller or disorder factor (${\sigma}^2$) obtained by EXAFS fitting for NiRuMnSn at Ni-, Ru- and Mn-edge.}

\begin{tabular}{|c|c|c|c|c|c|c|c|c|c|c|c|}
\hline\hline
\multicolumn{4}{|c|}{Ni edge} & \multicolumn{4}{c|}{Ru edge} & \multicolumn{4}{c|}{Mn edge}  \\   \hline
 Path & R  ({\AA})  & N & ${\sigma}^2$   & Path & R ({\AA}) & N & ${\sigma}^2$ &Path & R ({\AA}) & N & ${\sigma}^2$   \\ \hline
Ni-Mn&2.63${\pm}$0.01 &4 & 0.0106${\pm}$0.0017 & Ru-Mn &  2.57${\pm}$0.01 & 4 & 0.0210${\pm}$0.0048  & Mn-Ni  & 2.62${\pm}$0.03  &  5.4  & 0.0161${\pm}$0.0041                                           \\ \hline
Ni-Sn &  2.63${\pm}$0.01 & 4  & 0.0093${\pm}$0.0014& Ru-Sn &  2.57${\pm}$0.01 & 4 &  0.0095${\pm}$0.0030  & Mn-Ru & 2.60${\pm}$0.03  &  2.6  &  0.0057${\pm}$0.0030                                       \\ \hline
Ni-Ni &3.00${\pm}$0.03 &  1.7 & 0.0168${\pm}$0.0048 & Ru-Ni &  3.00${\pm}$0.03 & 3.5   &  0.0069${\pm}$0.0011  & Mn-Sn  & 3.06${\pm}$0.02 & 6 &0.0115${\pm}$0.0020                                         \\ \hline
Ni-Ru  & 3.00${\pm}$0.03 &  4.3 & 0.0135${\pm}$0.0048 &  Ru-Ru & 3.00${\pm}$0.03  & 2.5   &   0.0018${\pm}$0.0007  &Mn-Mn & 4.32${\pm}$0.05 &  12   &   0.0282${\pm}$0.0081                           \\ \hline
\end{tabular}
\label{Tab:EXAFS}
\end{table*}

The EXAFS fitted results are summarized in Table~\ref{Tab:EXAFS}. In the Fourier transformed EXAFS spectrum of Ni \textit{K}-edge (Fig.~\ref{Fig_6} (a)), the main peak near 2 {\AA} has contributions from Ni-Mn (2.63 {\AA}) and Ni-Sn (2.63 {\AA}) paths.
The small peak around 2.5 {\AA}  at Ni-edge spectrum includes contribution from Ni-Ni (3.00 {\AA}) and Ni-Ru  (3.00 {\AA}) paths.
For the case of Ru \textit{K}-edge, out of two main peaks observed in the Fourier transformed (FT) EXAFS spectrum (Fig.~\ref{Fig_6} (b)), the first peak near 2 {\AA} has contributions from Ru-Mn (2.57 {\AA}) and Ru-Sn (2.57 {\AA}) paths, whereas the second peak observed at 2.5 {\AA} has contribution from Ru-Ni  (3.00 {\AA}) and Ru-Ru (3.00 {\AA}) paths. In a similar way, for the case of Mn \textit{K}-edge, the first major peak near 2 {\AA} in the FT-EXAFS spectrum (Fig.~\ref{Fig_6} (c)) has contributions of Mn-Ni (2.62 {\AA}) and Mn-Ru (2.60{\AA}) paths.
The second peak in the Mn FT-EXAFS spectrum near 2.8 {\AA} has contribution from Mn-Sn (3.06 {\AA}) and Mn-Mn (4.32 {\AA}) paths.
Earlier X-ray diffraction results suggest a 50:50 swap disorder between Ni and Ru atoms.
It is worth to mention here that simultaneous fitting of all the edges (Ni-, Ru- and Mn-) in the EXAFS data  could only be possible assuming disordered structure.
Thus our EXAFS data are consistent with X-ray diffraction (Sec.\ref{sec:XRD}) and DFT calculations results (Table.~\ref{Enthalpy}).

\subsection{\label{sec:Magnetism}Magnetic properties}

\begin{figure}[h]
\centerline{\includegraphics[width=.48\textwidth]{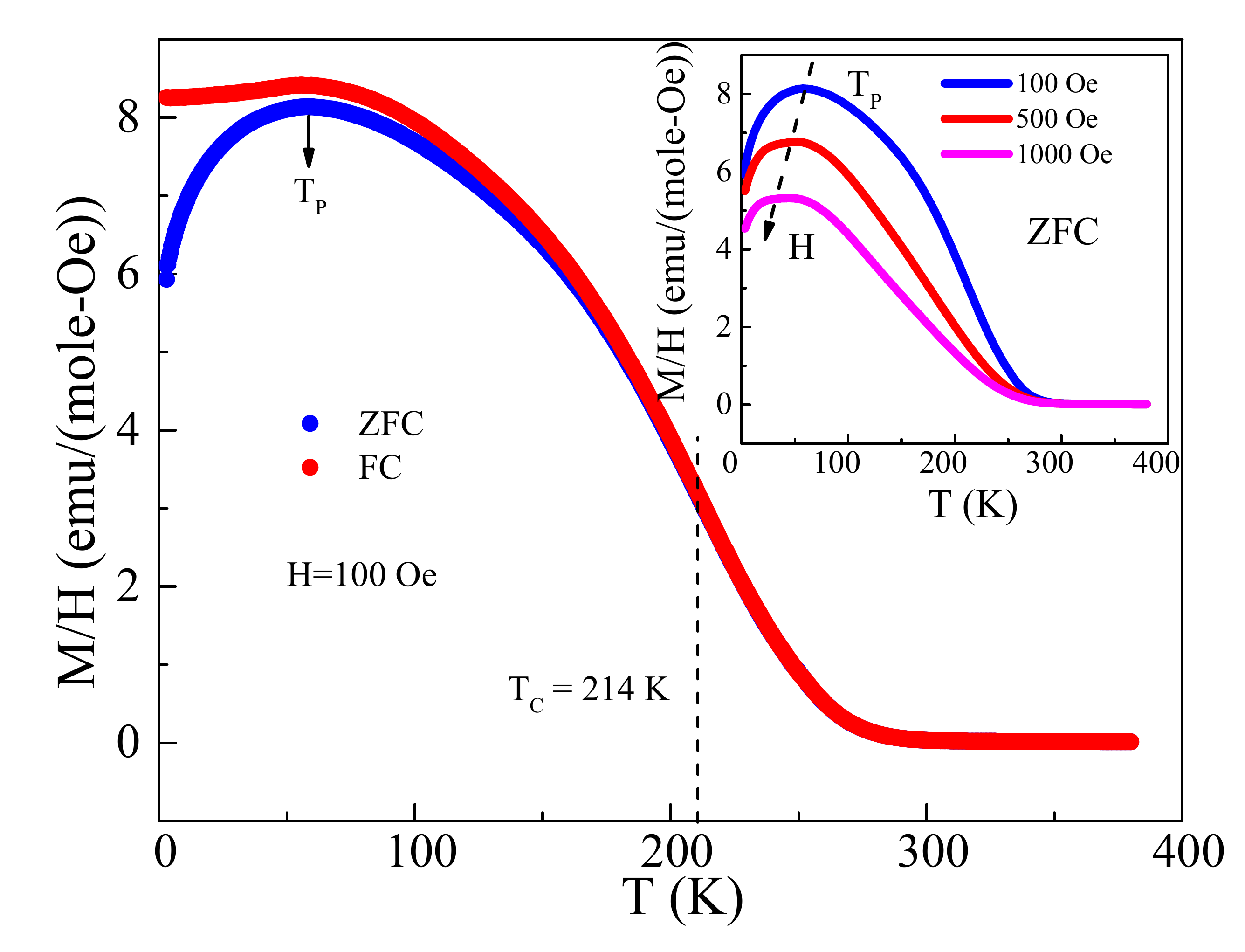}}
{\caption{Temperature dependence of  magnetization of NiRuMnSn measured under 100 Oe applied magnetic field under ZFC and FC configuration. Inset shows ZFC magnetization at different fields demonstrating the shift of peak temperature (T$_P$) with the application of field.}\label{Fig_7}}
\end{figure}

Fig.~\ref{Fig_7} represents the magnetization (M) \textit{vs.} temperature (T) of NiRuMnSn measured during zero-field-cooled (ZFC) heating and field-cooled (FC) cooling in 100 Oe magnetic field. The compound undergoes broad ferromagnetic$\longleftrightarrow$paramagnetic transitions at the Curie temperature (T$_{\rm C}$) centered near 214 K as determined from the minima of $d(M/H)/dT$. ZFC and FC data bifurcate below T$_{\rm C}$, and the ZFC curve shows a peak at T$_P$ around 60 K, below which magnetization begins to decrease with lowering temperature. In case of FC measurement, magnetization below T$_P$ is substantially enhanced compared to ZFC data, and the peak becomes nearly indistinguishable. T$_P$ shifts towards lower temperatures with the application of higher magnetic fields as shown in the inset of Fig.~\ref{Fig_7}. This type of behaviour is frequently observed in cases of anti-ferromagnetic (AFM) or spin-glass type states at low temperatures~\cite{samanta2018reentrant,kroder2019spin,zhang2014spin}. Based upon on the magnetization data three possible scenario could be envisaged : (i) a spin-reorientation or canting of existing ferromagnetic arrangement, (ii) because of structural disorder, only a fraction of the phase ordered ferromagnetically at T$_{\rm C}$ $\sim$ 214 K while the rest exhibits the magnetic transition at T$_P$ or (iii) magnetic inhomogeneity develops within a ferro-magnetically ordered phase, where multiple magnetic phases coexist below T$_P$. As the magnetic susceptibility can not distinguish between the above three possibilities, we have carried out isothermal magnetisation, neutron diffraction, neutron depolarization and ac susceptibility measurements for clarity.\\
\begin{figure}[h]
\centerline{\includegraphics[width=.48\textwidth]{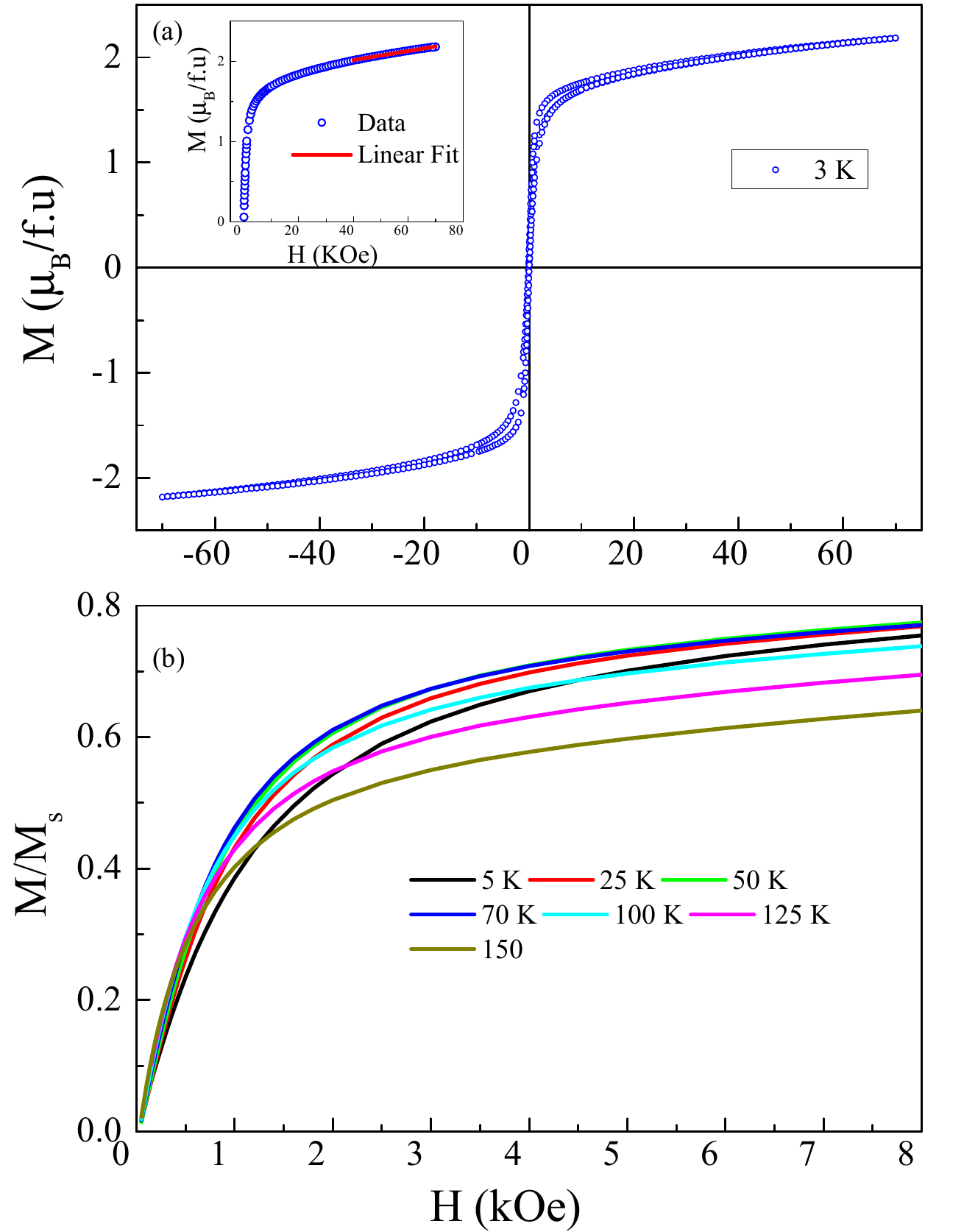}}
{\caption{(a) Isothermal magnetization (M(H)) of NiRuMnSn measured at 3 K\@. Inset shows linear fitting of the high field data. (b) Magnified view of the M(H) taken at different temperatures in the low field region, scaled with respect to the highest obtained value of M (defined as M$_S$) for a given temperature at 70 kOe.}\label{Fig_8}}
\end{figure}

Fig.~\ref{Fig_8}(a) illustrates the isothermal magnetization, $M(H)$, of the title material measured at 3 K after zero-field cooling. Despite a ferromagnetic-like character, M(H) remains far from saturation even at 70 kOe, apparently being a superposition of (dominant) soft-ferromagnetic and (weaker) linear contributions, indicating presence of an AFM or a paramagnetic phase, in addition to the FM phase (Fig.~\ref{Fig_8}(a)). Moreover, the magnetic moment under application of 70 kOe at 3 K is found to be  of  2.17 $\mu_B$/f.u, which appears to be considerably smaller in comparison to 5 $\mu_B$/f.u. expected from the Slater-Pauling (S-P) rule~\cite{galanakis2002slater,ozdougan2013slater}. The spontaneous magnetization, which is defined as the net magnetisation of the system in the absence of external magnetic field, is estimated from the linear extrapolation of saturation magnetisation to $H = 0$\,Oe and found to be 1.77\,$\mu_B$/f.u. (inset of Fig.~\ref{Fig_8}(a)).Comparing NiRuMnSn and NiFeMnSn, it is clear that introduction of the 4\textit{d}-element Ru dramatically alters magnetic properties of a purely 3\textit{d}-element based NiFeMnSn~\cite{mukadam2016quantification}. Thus the Curie temperature is reduced from 405 K in NiFeMnSn to 214 K in NiRuMnSn, and the saturation magnetization at low temperatures is suppressed from 4.18 $\mu_B$/f.u. in NiFeMnSn to 2.17  $\mu_B$/f.u. in NiRuMnSn. This is quite unusual as in most of the known Heusler alloys (\textit{e.g.} CoFeMnSi~\cite{bainsla2015spin}, CoRhMnGe~\cite{rani2017structural}, CoFeMnGe~\cite{bainsla2014high} and CoRuMnSi~\cite{venkateswara2020half}), the substitutions of Fe with Ru or Rh do not affect the magnetism in such a drastic way, and importantly, compounds containing Co in \textit{X} site closely obey the S-P rule. Interestingly, a closer investigation of the magnetic isotherms (Fig.~\ref{Fig_8}(b)) reveals the relative magnetisation values measured in the low field region and at temperatures below T$_P$ to be smaller than that of the data measured above T$_P$. For a purely ferromagnetic system, such behaviour is not expected and thus suggest that contribution of ferromagnetic phase got weakened due to the disturbance of the ferromagnetic spin structure by spin-canting or reorientation or development of spin/cluster glass behaviour. Since neutron diffraction measurements can detect the change in magnetic spin structure, we have carried them out at different temperatures in the range 1.5-300 K.

\subsection{\label{sec:Neutron}Neutron diffraction and neutron depolarization}

\begin{figure}[h]
\centerline{\includegraphics[width=.48\textwidth]{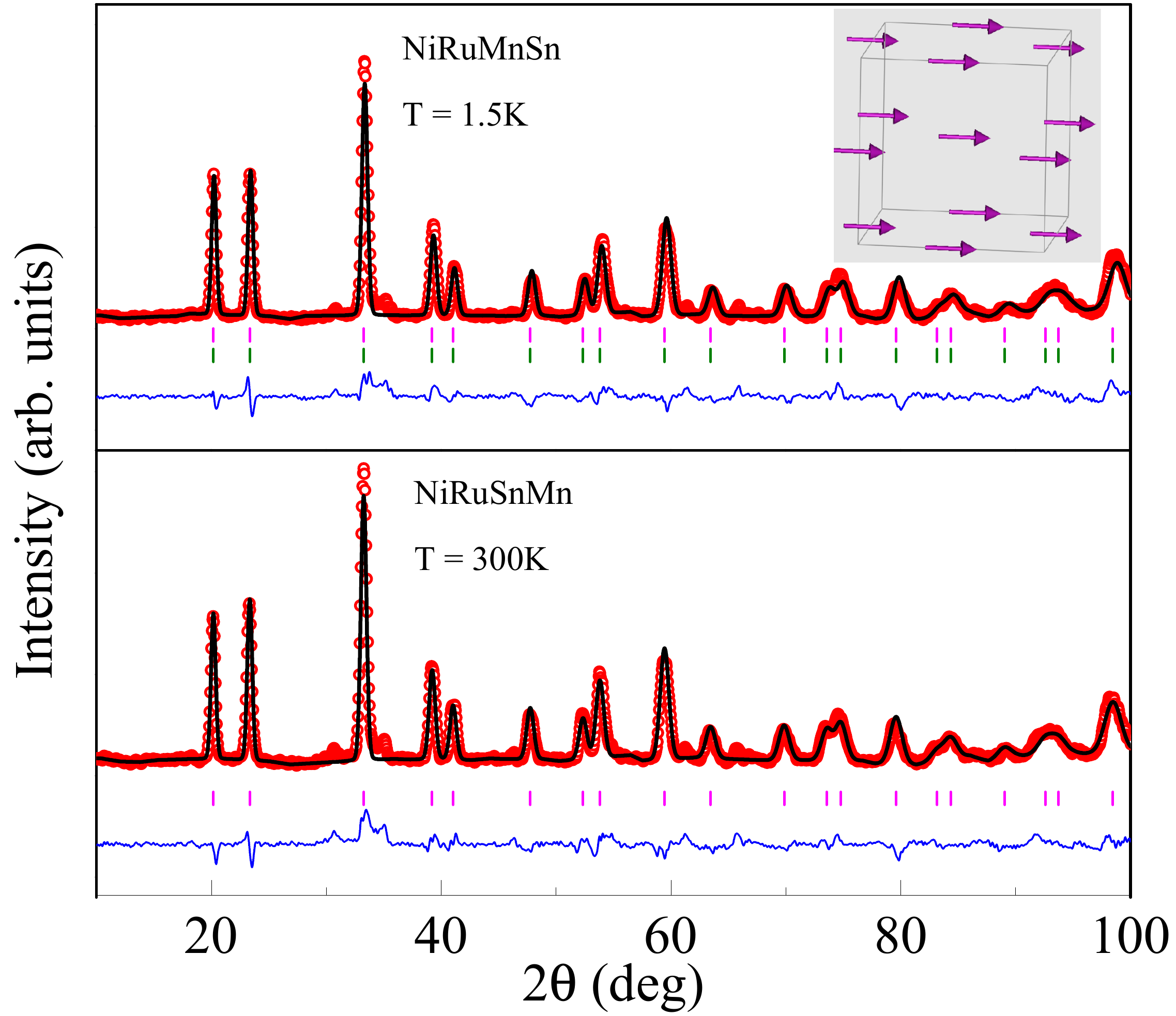}}
{\caption{Rietveld refinement of the neutron diffraction pattern of NiRuMnSn taken at 1.5 K and 300 K. Inset of the upper figure depicts the magnetic structure of NiRuMnSn, where spins are developed on the Mn atom.}\label{Fig_9}}
\end{figure}

Neutron diffraction measurement is not only helpful in understanding the magnetic structure, but can also be utilized as a complementary tool to XRD for detailed structural characterisation of a system, particularly when multiple atoms with similar atomic numbers are involved. The scattering lengths of Ni, Ru, Mn, and Sn are 1.030$\times$10$^{-12}$ cm, 0.7030$\times$10$^{-12}$, -0.3730$\times$10$^{-12}$ and 0.6225$\times$10$^{-12}$, respectively, making it possible to confirm the site occupancies earlier postulated from the analysis of the X-ray diffraction data. Fig.~\ref{Fig_9} shows the Rietveld-refined neutron diffraction patterns recorded at 300 K and 1.5 K. The data at 300 K are fitted in $Fm\bar{3}m$ space group with atoms occupying the following positions: Sn in 4\textit{a} (0,0,0); Mn in 4\textit{b} (0.5, 0.5, 0.5) with full occupancy; Ni/Ru in 4\textit{c} (0.25, 0.25, 0.25) and 4\textit{d} (0.75, 0.75, 0.75) with 50/50 occupancy. A fully random distribution of Ni and Ru between 4\textit{c} and 4\textit{d} sites yields the best fit, in agreement with the analysis of XRD and EXAFS data.

At 1.5 K, weak enhancements in the intensities of (111) and (200) reflections are observed (Fig.~\ref{Fig_9}). Absence of additional reflections rules out the presence of long range antiferromagnetic (AFM) ordering. The enhanced intensities of the fundamental reflections at 1.5 K indicate FM arrangement of the magnetic moments, which was modelled assuming Mn as the only magnetic species (consistent with DFT calculations, Sec.~\ref{sec:Electronic_Structure_Ordered}) in \textit{F-1} space group.
The refined total magnetic moment of Mn at 1.5 K is 1.69 $\mu_B$/f.u., which is consistent, within error, with 1.77 $\mu_B$/f.u. spontaneous magnetic moment estimated from the isothermal magnetization measurements at 3 K (inset of Fig.~\ref{Fig_8}(a)).
Analysis of ND data could not detect any moment on Ni-site, suggesting the Ni-moment, if any, in this compound  should be below the limit of resolution of the experiment. This is expected because the Ni and Ru atoms at 4\textit{c} and 4\textit{d} sites are randomly distributed, which strongly hinders the possibility of long range ordering of the Ni-spins. The absence of any change in magnetic structure before and after T$_P$ excludes the possibility of any spin canting/reorientation in the system. Thus, the anomaly in magnetic susceptibility at T$_P$ can only be attributed to the development of spin/cluster glass phase. Combining the observation of lower magnetic moment in the low field region for T $<$ T$_P$, and the conservation of the ferromagnetic spin structure at lowest temperature suggested by the neutron diffraction measurement, the reduction of M(T) in both ZFC and FC configurations is likely due to the development of a glassy phase at the partial expense of ferromagnetically ordered spins.

Neutron depolarisation experiment is a very effective tool to study the growth and development of ferromagnetic domains. Accordingly we have carried out the neutron depolarisation measurement in the temperature range 2$-$300 K, covering both the T$_{\rm C}$ and T$_P$. In the neutron depolarisation experiment, polarization vector of a polarized neutron beam is examined after the beam passes through a magnetic medium. As the magnetic inhomogeneities in the medium influence the polarization vector during transmission, the mean magnetic induction causes a net rotation of the polarization vector. In presence of spontaneous magnetization, the local magnetic induction results in an effective  rotation of the polarization vector, which is known as depolarization. This characteristic can be used to distinguish a purely ferromagnetic system from any other kind of magnetic ordering. The frequency of spin fluctuations in the paramagnetic state is too high for neutron polarization to be affected and no depolarization is observed~\cite{das2003neutron,das1999neutron}. In a spin-glass system, as the spins are frozen randomly, the random spatial fluctuations of the effective field can not influence the polarization vector either. Depolarization is also not expected in AFM materials since net magnetization is zero. In contrast, in ferromagnets, the neutron spins experience torque around the magnetization axes as they travel through equally magnetized domains resulting in the loss of polarization~\cite{halpern1941passage,mitsuda1992neutron,samanta2018reentrant}. Thus, by using the neutron depolarization experiment, one can estimate the mean orientation of these fluctuations (magnetic texture), the mean magnetization, and the magnetic correlation length along the neutron route.

\begin{figure}[h]
\centerline{\includegraphics[width=.48\textwidth]{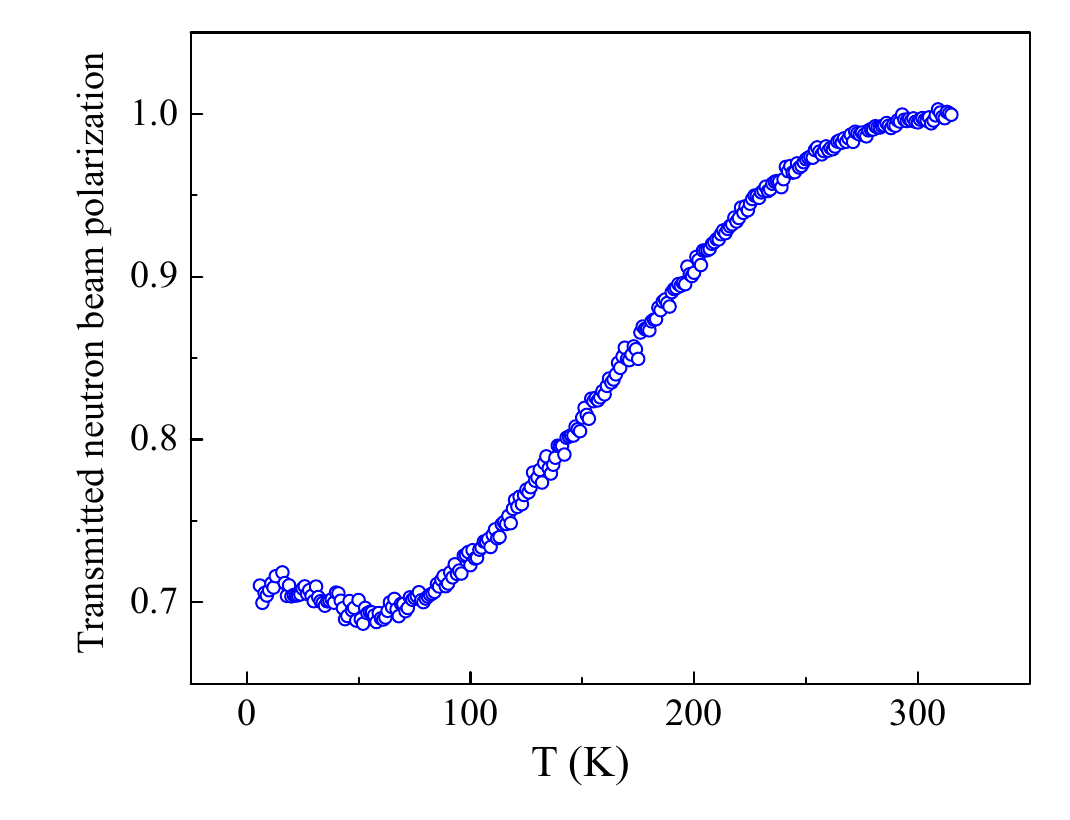}}
{\caption{Temperature dependence of the transmitted neutron polarization under H = 50 Oe.}\label{Fig_10}}
\end{figure}

For present NiRuMnSn sample, neutron depolarization is measured in the temperature range 5$-$300 K in the presence of a magnetic field of 50 Oe in zero field cooled (ZFC) mode (Fig.~\ref{Fig_10}). As the temperature is lowered from room temperature, the polarization does not exhibit any significant changes in the paramagnetic region. As one approaches towards and moves further below the Curie temperature, the development and subsequent gradual increase of domain size is reflected in the decreasing trend of the neutron beam polarization signal, as  expected in a ferromagnetic material. The depolarization signal shows a minimum at $\sim$60 K below which the polarization signal starts to grow again. The minimum in polarization around 60 K and increasing nature of polarization for T $<$60 K can be explained with the decrease in the coherence length of the ferromagnetic interaction due to the decrease of the ferromagnetic domain sizes. The size of the magnetic domain can be estimated using the expression~\cite{halpern1941passage,das1999neutron}

\begin{equation}
P_f = P_iexp\Big[-\alpha({\frac{d}{\Delta}}){<\phi_\delta>}^2\Big]
\end{equation}
where $P_i$ and $P_f$ are the initial and final transmitted beam polarization, ${\alpha}$ is a dimensionless parameter equal to 1/3, \textit{d} is the sample thickness which is 7\,mm in the studied sample, ${\Delta}$ is the domain size and the precession angle ${\phi_\delta}$ =${(4.63\times10^{-10} G^{-1}{\AA}^{-2}){\lambda}B\Delta}$.
The domain magnetization ($B$) is obtained from the bulk magnetization.
The domain size thus estimated to be 0.2\,${\mu}$m at 5\,K.
The change in depolarisation values below 60\,K signifies the reduction in domain sizes, \textit{i.e.}, enhancement of number of magnetic spins in the domain boundaries by deviating from the domain magnetisation direction.
To understand whether such reduction in domain size causes any magnetic phase coexistence, we have carried out the ac susceptibility measurements.

\subsection{\label{sec:Ac}ac susceptibility}

\begin{figure}[h]
\centerline{\includegraphics[width=.48\textwidth]{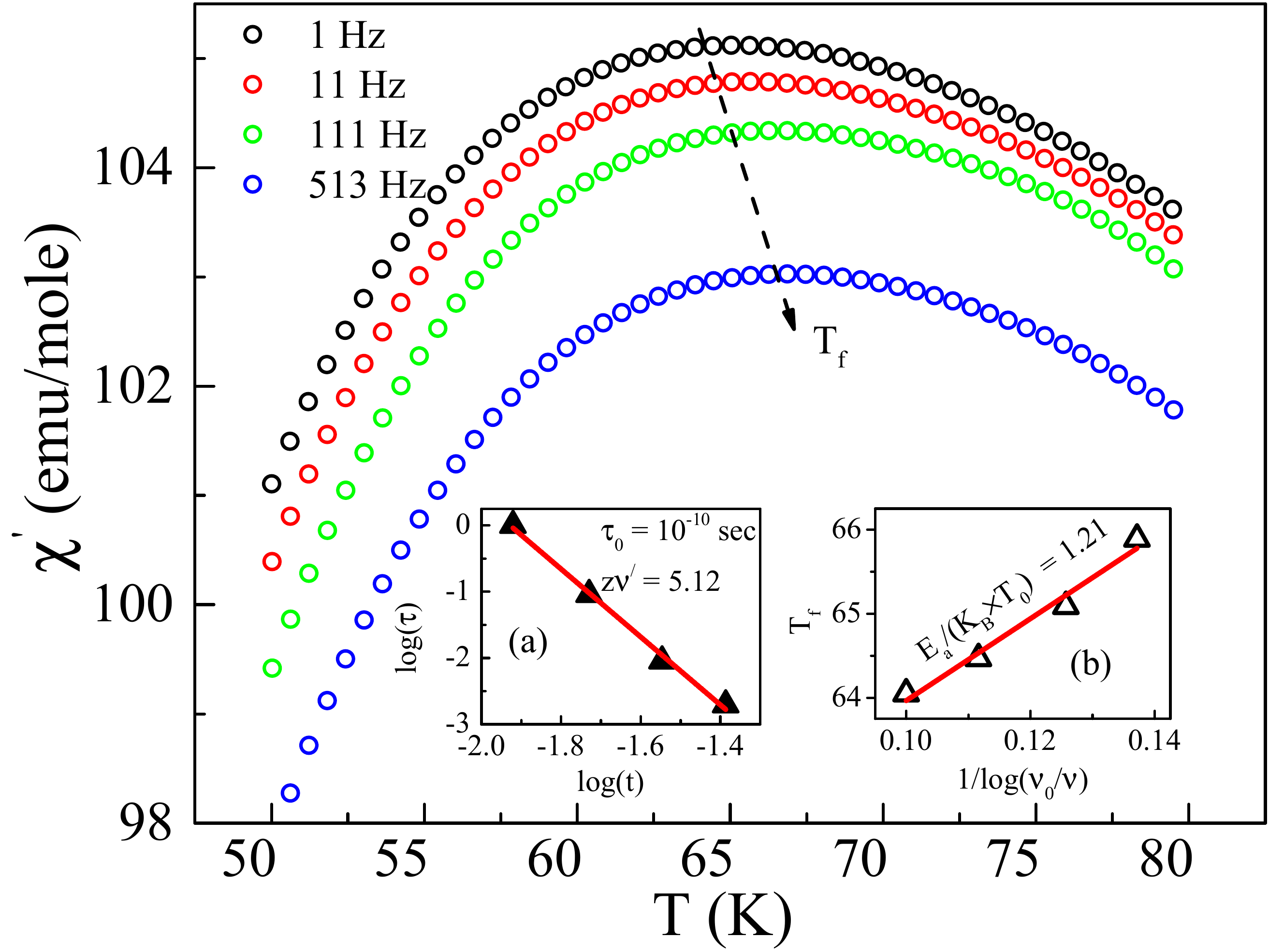}}
{\caption{Temperature dependence of ${\chi}^{\prime}$ (T) of NiRuMnSn taken at different frequencies. Inset (a) shows the variation of log($\tau$) with log(t). Inset (b) represents the plot of T$_f$ \textit{vs.} 1/$\log\frac{\nu_0}{\nu}$.}\label{Fig_11}}
\end{figure}

Fig.~\ref{Fig_11} represents the temperature variations of the real part of the ac susceptibility (${\chi}^{\prime}$) performed under low ac field of 5 Oe  and different frequencies.
${\chi}^{\prime}$ ($T$) data shows well-rounded peak near 65\,K which indicates and the frequency dependent character of the peak  of $\chi'$(T) suggests the magnetic anomaly is of spin-glass type.
In-order to determine the nature of the spin-glass state, \textit{i.e.}, whether it is a canonical or cluster-glass type, we have carried out a detail analysis of the nature of temperature variation of the peak in ${\chi}^{\prime}$ ($T$).

The relative shift in freezing temperature per decade of frequency in a typical glassy system is commonly represented as~\cite{mydosh1993spin}.

\begin{eqnarray}
\delta T_{f} =\frac{\Delta T_{f}}{T_{f}\Delta (\log_{10} \nu)}
\end{eqnarray}
where, $T_f$ is the freezing temperature and $\nu$ is the frequency.
The value of $\delta T_{f}$ have been found to be $\sim$0.001 for canonical spin glasses~\cite{mondal2020identification}, $\sim$0.01 for spin-cluster glass compounds~\cite{pakhira2016large,pakhira2018magnetic,chakraborty2022ground}, and $\sim$0.1 for numerous known superparamagnetic systems~\cite{pakhira2016large}.
In the studied compound, $\delta T_{f}$  found to be 0.008 which lies in the  borderline of canonical spin glass and cluster-glass regime.
$\delta T_{f}$ for a spin glass state also follows a frequency dependence given by~\cite{mydosh1993spin,mori2003dynamical}
\begin{eqnarray}
\tau = \tau_0\left(\frac{T_f-T_{SG}}{T_{SG}}\right)^{-z\nu^{\prime}}\label{eqn:dynamical scaling}
\end{eqnarray}
where $\tau$ is the relaxation time associated with the measured frequency ($\tau$=1/{$\nu$}), $\tau_0$ is the single-flip relaxation time, $T_{SG}$ is the spin-glass temperature for $\nu$=0, and z$\nu^{\prime}$ is the dynamical critical exponent.
The value of z$\nu^{\prime}$ typically lies between 4--12 for spin glass state.
The value of $\tau_0$ for canonical spin glasses lies in the region of $10^{-13}$ --$10^{-12}$ second, but for a spin cluster glass system it lies mostly in the range of $10^{-11}$--$10^{-4}$ second.
On the other hand, superparamagnetic states are associated with larger values of  $\tau_0$~\cite{lago2012three,pakhira2016large}.
For NiRuMnSn, the value of z$\nu^{\prime}$ found to be 5.1 which is in the range expected for spin glass state formation while the estimated  $\tau_0$ $\approx$ $10^{-10}$ second also lies in the border range of canonical and cluster-glass state. \\

Another dynamical scaling law, known as the empirical Vogel-Fulcher relation, can be used to simulate spin dynamics in glassy systems around the freezing temperature.
The frequency dependence of freezing temperature can be stated as~\cite{mydosh1993spin,souletie1985critical}
\begin{eqnarray}
\nu = \nu_{0} \exp \left[-\frac{E_a}{k_B(T_f-T_0)}\right]\label{eqn:Vogel-Fulcher}
\end{eqnarray}
where $\nu_{0}$ is known as the characteristic attempt frequency, $T_{0}$ is called Vogel-Fulcher temperature and $E_a$ is the activation energy.
From the $T_f$ \textit{vs.} 1/$\log\frac{\nu_0}{\nu}$  plot for NiRuMnSn, the fitted values are found to be $E_a$/K$_B$  $\approx$ 59\,K and $T_0$ $\approx$ 48.7\,K. For canonical spin glass state, the value of $\frac {E_a} {K_{B}T_{0}}$ generally comes out to be close to 1 whereas for cluster glass type of system it assumes a relatively larger value.
In the studied compound, the value of $\frac {E_a} {K_{B}T_{0}}$ is estimated to be $\sim$1.21 which is in the canonical spin glass regime.
Combining neutron polarization and ac-susceptibility results, it can be said that the domain boundary spins lost its FM nature below 60\,K (discussed in Sec.~\ref{sec:Neutron}) and are frozen in space giving rise to spin glass state.
NiRuMnSn can be viewed as a reentrant spin glass system where the glassy state formed below the $T_{\rm C}$.

\section{Conclusion}
We have synthesized a new 4\textit{d}-based quaternary Heusler alloy NiRuMnSn through arc melting technique.
The combined studies utilizing XRD, EXAFS, and neutron diffraction reveal that the compound crystallizes in a partially disordered structure, where Ni and Ru statistically mix in the 4\textit{c} and 4\textit{d} wyckoff  positions. This experimental observation  is also supported by the theoretical estimation of lower formation ground state energy in the disordered structure, containing mixed Ni and Ru atoms in same plane, over the fully ordered one.
The compound exhibits a ferromagnetic to paramagnetic transition near 214\,K and the saturation moment at 3\,K is estimated to be 2.17\,$\mu_B$/f.u, which is much lower than the expected Slater-Pauli value of 5 $\mu_B$/f.u.
Neutron diffraction experiments further reveal that the magnetic moment is carried only by the Mn atoms, in agreement with the DFT calculations.
Neutron depolarization along with ac susceptibility results confirms that the gradual growth of the ferromagnetic domains with lowering of temperature is hindered below 60 K due to the development of reentrant spin--glass state. A discrepancy between the theoretically estimated moment with the experimentally observed value has been argued to be due to the Ni/Ru site disorder along with a development of spin-glass phase at low temperature. It would be interesting to note that among about a thousand of Heusler alloys reported so far, only a handful few have been reported to show spin glass behaviour, and, thus, NiRuMnSn is a very rare system among the equiatomic quaternary Heusler alloys. Although it is generally believed that isoelectronic substitution of 3\textit{d} metal with 4\textit{d} element that has  larger atomic radii is likely to improve the structural order as well as enhance spin-polarisation in Heusler alloys, our theoretical and experimental results proves that while the structural order indeed improves in NiRuMnSn ($L2_1$-type) in comparison to NiFeMnSn ($D0_3$-type), the nonmagnetic character of isoelectronic Ru (in comparison to Fe) resulted in the much lower spin polarisation value even in ordered structure. \\

\centerline{\textbf{Acknowledgement}}

We would like to dedicate this paper in memory of senior co-author: Late Prof. Vitalij K. Pecharsky,  he actively participated in this work  and recently passed away before submission of the paper.  S.G and S.C would like to sincerely acknowledge SINP, India and UGC, India, respectively, for their fellowship. DFT calculations were performed using HPC resources from GENCI-CINES (Grant 2021-A0100906175). Work at the Ames Laboratory was supported by the Department of Energy- Basic Energy Sciences, Materials Sciences and Engineering Division, under Contract No. DE-AC02-07CH11358.
\normalem
\bibliographystyle{apsrev4-2}
\bibliography{reference}

\begin{thebibliography}{65}%
\makeatletter
\providecommand \@ifxundefined [1]{%
 \@ifx{#1\undefined}
}%
\providecommand \@ifnum [1]{%
 \ifnum #1\expandafter \@firstoftwo
 \else \expandafter \@secondoftwo
 \fi
}%
\providecommand \@ifx [1]{%
 \ifx #1\expandafter \@firstoftwo
 \else \expandafter \@secondoftwo
 \fi
}%
\providecommand \natexlab [1]{#1}%
\providecommand \enquote  [1]{``#1''}%
\providecommand \bibnamefont  [1]{#1}%
\providecommand \bibfnamefont [1]{#1}%
\providecommand \citenamefont [1]{#1}%
\providecommand \href@noop [0]{\@secondoftwo}%
\providecommand \href [0]{\begingroup \@sanitize@url \@href}%
\providecommand \@href[1]{\@@startlink{#1}\@@href}%
\providecommand \@@href[1]{\endgroup#1\@@endlink}%
\providecommand \@sanitize@url [0]{\catcode `\\12\catcode `\$12\catcode
  `\&12\catcode `\#12\catcode `\^12\catcode `\_12\catcode `\%12\relax}%
\providecommand \@@startlink[1]{}%
\providecommand \@@endlink[0]{}%
\providecommand \url  [0]{\begingroup\@sanitize@url \@url }%
\providecommand \@url [1]{\endgroup\@href {#1}{\urlprefix }}%
\providecommand \urlprefix  [0]{URL }%
\providecommand \Eprint [0]{\href }%
\providecommand \doibase [0]{https://doi.org/}%
\providecommand \selectlanguage [0]{\@gobble}%
\providecommand \bibinfo  [0]{\@secondoftwo}%
\providecommand \bibfield  [0]{\@secondoftwo}%
\providecommand \translation [1]{[#1]}%
\providecommand \BibitemOpen [0]{}%
\providecommand \bibitemStop [0]{}%
\providecommand \bibitemNoStop [0]{.\EOS\space}%
\providecommand \EOS [0]{\spacefactor3000\relax}%
\providecommand \BibitemShut  [1]{\csname bibitem#1\endcsname}%
\let\auto@bib@innerbib\@empty
\bibitem [{\citenamefont {De~Groot}\ \emph {et~al.}(1983)\citenamefont
  {De~Groot}, \citenamefont {Mueller}, \citenamefont {Van~Engen},\ and\
  \citenamefont {Buschow}}]{de1983new}%
  \BibitemOpen
  \bibfield  {author} {\bibinfo {author} {\bibfnamefont {R.~A.}\ \bibnamefont
  {De~Groot}}, \bibinfo {author} {\bibfnamefont {F.~M.}\ \bibnamefont
  {Mueller}}, \bibinfo {author} {\bibfnamefont {P.~G.}\ \bibnamefont
  {Van~Engen}},\ and\ \bibinfo {author} {\bibfnamefont {K.~H.~J.}\ \bibnamefont
  {Buschow}},\ }\href@noop {} {\bibfield  {journal} {\bibinfo  {journal} {Phys.
  Rev. Lett.}\ }\textbf {\bibinfo {volume} {50}},\ \bibinfo {pages} {2024}
  (\bibinfo {year} {1983})}\BibitemShut {NoStop}%
\bibitem [{\citenamefont {Ma{\~n}osa}\ \emph {et~al.}(2010)\citenamefont
  {Ma{\~n}osa}, \citenamefont {Gonz{\'a}lez-Alonso}, \citenamefont {Planes},
  \citenamefont {Bonnot}, \citenamefont {Barrio}, \citenamefont {Tamarit},
  \citenamefont {Aksoy},\ and\ \citenamefont {Acet}}]{manosa2010giant}%
  \BibitemOpen
  \bibfield  {author} {\bibinfo {author} {\bibfnamefont {L.}~\bibnamefont
  {Ma{\~n}osa}}, \bibinfo {author} {\bibfnamefont {D.}~\bibnamefont
  {Gonz{\'a}lez-Alonso}}, \bibinfo {author} {\bibfnamefont {A.}~\bibnamefont
  {Planes}}, \bibinfo {author} {\bibfnamefont {E.}~\bibnamefont {Bonnot}},
  \bibinfo {author} {\bibfnamefont {M.}~\bibnamefont {Barrio}}, \bibinfo
  {author} {\bibfnamefont {J.-L.}\ \bibnamefont {Tamarit}}, \bibinfo {author}
  {\bibfnamefont {S.}~\bibnamefont {Aksoy}},\ and\ \bibinfo {author}
  {\bibfnamefont {M.}~\bibnamefont {Acet}},\ }\href@noop {} {\bibfield
  {journal} {\bibinfo  {journal} {Nat. Mater.}\ }\textbf {\bibinfo {volume}
  {9}},\ \bibinfo {pages} {478} (\bibinfo {year} {2010})}\BibitemShut {NoStop}%
\bibitem [{\citenamefont {Nayak}\ \emph {et~al.}(2017)\citenamefont {Nayak},
  \citenamefont {Kumar}, \citenamefont {Ma}, \citenamefont {Werner},
  \citenamefont {Pippel}, \citenamefont {Sahoo}, \citenamefont {Damay},
  \citenamefont {R{\"o}{\ss}ler}, \citenamefont {Felser},\ and\ \citenamefont
  {Parkin}}]{nayak2017magnetic}%
  \BibitemOpen
  \bibfield  {author} {\bibinfo {author} {\bibfnamefont {A.~K.}\ \bibnamefont
  {Nayak}}, \bibinfo {author} {\bibfnamefont {V.}~\bibnamefont {Kumar}},
  \bibinfo {author} {\bibfnamefont {T.}~\bibnamefont {Ma}}, \bibinfo {author}
  {\bibfnamefont {P.}~\bibnamefont {Werner}}, \bibinfo {author} {\bibfnamefont
  {E.}~\bibnamefont {Pippel}}, \bibinfo {author} {\bibfnamefont
  {R.}~\bibnamefont {Sahoo}}, \bibinfo {author} {\bibfnamefont
  {F.}~\bibnamefont {Damay}}, \bibinfo {author} {\bibfnamefont {U.~K.}\
  \bibnamefont {R{\"o}{\ss}ler}}, \bibinfo {author} {\bibfnamefont
  {C.}~\bibnamefont {Felser}},\ and\ \bibinfo {author} {\bibfnamefont {S.~S.}\
  \bibnamefont {Parkin}},\ }\href@noop {} {\bibfield  {journal} {\bibinfo
  {journal} {Nature}\ }\textbf {\bibinfo {volume} {548}},\ \bibinfo {pages}
  {561} (\bibinfo {year} {2017})}\BibitemShut {NoStop}%
\bibitem [{\citenamefont {Manna}\ \emph {et~al.}(2018)\citenamefont {Manna},
  \citenamefont {Sun}, \citenamefont {Muechler}, \citenamefont {K{\"u}bler},\
  and\ \citenamefont {Felser}}]{manna2018heusler}%
  \BibitemOpen
  \bibfield  {author} {\bibinfo {author} {\bibfnamefont {K.}~\bibnamefont
  {Manna}}, \bibinfo {author} {\bibfnamefont {Y.}~\bibnamefont {Sun}}, \bibinfo
  {author} {\bibfnamefont {L.}~\bibnamefont {Muechler}}, \bibinfo {author}
  {\bibfnamefont {J.}~\bibnamefont {K{\"u}bler}},\ and\ \bibinfo {author}
  {\bibfnamefont {C.}~\bibnamefont {Felser}},\ }\href@noop {} {\bibfield
  {journal} {\bibinfo  {journal} {Nat. Rev. Mater.}\ }\textbf {\bibinfo
  {volume} {3}},\ \bibinfo {pages} {244} (\bibinfo {year} {2018})}\BibitemShut
  {NoStop}%
\bibitem [{\citenamefont {Hinterleitner}\ \emph {et~al.}(2019)\citenamefont
  {Hinterleitner}, \citenamefont {Knapp}, \citenamefont {Poneder},
  \citenamefont {Shi}, \citenamefont {M{\"u}ller}, \citenamefont {Eguchi},
  \citenamefont {Eisenmenger-Sittner}, \citenamefont {St{\"o}ger-Pollach},
  \citenamefont {Kakefuda}, \citenamefont {Kawamoto} \emph
  {et~al.}}]{hinterleitner2019thermoelectric}%
  \BibitemOpen
  \bibfield  {author} {\bibinfo {author} {\bibfnamefont {B.}~\bibnamefont
  {Hinterleitner}}, \bibinfo {author} {\bibfnamefont {I.}~\bibnamefont
  {Knapp}}, \bibinfo {author} {\bibfnamefont {M.}~\bibnamefont {Poneder}},
  \bibinfo {author} {\bibfnamefont {Y.}~\bibnamefont {Shi}}, \bibinfo {author}
  {\bibfnamefont {H.}~\bibnamefont {M{\"u}ller}}, \bibinfo {author}
  {\bibfnamefont {G.}~\bibnamefont {Eguchi}}, \bibinfo {author} {\bibfnamefont
  {C.}~\bibnamefont {Eisenmenger-Sittner}}, \bibinfo {author} {\bibfnamefont
  {M.}~\bibnamefont {St{\"o}ger-Pollach}}, \bibinfo {author} {\bibfnamefont
  {Y.}~\bibnamefont {Kakefuda}}, \bibinfo {author} {\bibfnamefont
  {N.}~\bibnamefont {Kawamoto}}, \emph {et~al.},\ }\href@noop {} {\bibfield
  {journal} {\bibinfo  {journal} {Nature}\ }\textbf {\bibinfo {volume} {576}},\
  \bibinfo {pages} {85} (\bibinfo {year} {2019})}\BibitemShut {NoStop}%
\bibitem [{\citenamefont {Mondal}\ \emph {et~al.}(2018)\citenamefont {Mondal},
  \citenamefont {Mazumdar}, \citenamefont {Ranganathan}, \citenamefont
  {Alleno}, \citenamefont {Sreeparvathy}, \citenamefont {Kanchana},\ and\
  \citenamefont {Vaitheeswaran}}]{mondal2018ferromagnetically}%
  \BibitemOpen
  \bibfield  {author} {\bibinfo {author} {\bibfnamefont {S.}~\bibnamefont
  {Mondal}}, \bibinfo {author} {\bibfnamefont {C.}~\bibnamefont {Mazumdar}},
  \bibinfo {author} {\bibfnamefont {R.}~\bibnamefont {Ranganathan}}, \bibinfo
  {author} {\bibfnamefont {E.}~\bibnamefont {Alleno}}, \bibinfo {author}
  {\bibfnamefont {P.}~\bibnamefont {Sreeparvathy}}, \bibinfo {author}
  {\bibfnamefont {V.}~\bibnamefont {Kanchana}},\ and\ \bibinfo {author}
  {\bibfnamefont {G.}~\bibnamefont {Vaitheeswaran}},\ }\href@noop {} {\bibfield
   {journal} {\bibinfo  {journal} {Physical Review B}\ }\textbf {\bibinfo
  {volume} {98}},\ \bibinfo {pages} {205130} (\bibinfo {year}
  {2018})}\BibitemShut {NoStop}%
\bibitem [{\citenamefont {Liu}\ \emph {et~al.}(2012)\citenamefont {Liu},
  \citenamefont {Gottschall}, \citenamefont {Skokov}, \citenamefont {Moore},\
  and\ \citenamefont {Gutfleisch}}]{liu2012giant}%
  \BibitemOpen
  \bibfield  {author} {\bibinfo {author} {\bibfnamefont {J.}~\bibnamefont
  {Liu}}, \bibinfo {author} {\bibfnamefont {T.}~\bibnamefont {Gottschall}},
  \bibinfo {author} {\bibfnamefont {K.~P.}\ \bibnamefont {Skokov}}, \bibinfo
  {author} {\bibfnamefont {J.~D.}\ \bibnamefont {Moore}},\ and\ \bibinfo
  {author} {\bibfnamefont {O.}~\bibnamefont {Gutfleisch}},\ }\href@noop {}
  {\bibfield  {journal} {\bibinfo  {journal} {Nature materials}\ }\textbf
  {\bibinfo {volume} {11}},\ \bibinfo {pages} {620} (\bibinfo {year}
  {2012})}\BibitemShut {NoStop}%
\bibitem [{\citenamefont {Graf}\ \emph {et~al.}(2011)\citenamefont {Graf},
  \citenamefont {Felser},\ and\ \citenamefont {Parkin}}]{graf2011simple}%
  \BibitemOpen
  \bibfield  {author} {\bibinfo {author} {\bibfnamefont {T.}~\bibnamefont
  {Graf}}, \bibinfo {author} {\bibfnamefont {C.}~\bibnamefont {Felser}},\ and\
  \bibinfo {author} {\bibfnamefont {S.~S.}\ \bibnamefont {Parkin}},\
  }\href@noop {} {\bibfield  {journal} {\bibinfo  {journal} {Prog. Solid. State
  Ch.}\ }\textbf {\bibinfo {volume} {39}},\ \bibinfo {pages} {1} (\bibinfo
  {year} {2011})}\BibitemShut {NoStop}%
\bibitem [{\citenamefont {Telling}\ \emph {et~al.}(2008)\citenamefont
  {Telling}, \citenamefont {Keatley}, \citenamefont {van~der Laan},
  \citenamefont {Hicken}, \citenamefont {Arenholz}, \citenamefont {Sakuraba},
  \citenamefont {Oogane}, \citenamefont {Ando}, \citenamefont {Takanashi},
  \citenamefont {Sakuma} \emph {et~al.}}]{telling2008evidence}%
  \BibitemOpen
  \bibfield  {author} {\bibinfo {author} {\bibfnamefont {N.}~\bibnamefont
  {Telling}}, \bibinfo {author} {\bibfnamefont {P.~S.}\ \bibnamefont
  {Keatley}}, \bibinfo {author} {\bibfnamefont {G.}~\bibnamefont {van~der
  Laan}}, \bibinfo {author} {\bibfnamefont {R.}~\bibnamefont {Hicken}},
  \bibinfo {author} {\bibfnamefont {E.}~\bibnamefont {Arenholz}}, \bibinfo
  {author} {\bibfnamefont {Y.}~\bibnamefont {Sakuraba}}, \bibinfo {author}
  {\bibfnamefont {M.}~\bibnamefont {Oogane}}, \bibinfo {author} {\bibfnamefont
  {Y.}~\bibnamefont {Ando}}, \bibinfo {author} {\bibfnamefont {K.}~\bibnamefont
  {Takanashi}}, \bibinfo {author} {\bibfnamefont {A.}~\bibnamefont {Sakuma}},
  \emph {et~al.},\ }\href@noop {} {\bibfield  {journal} {\bibinfo  {journal}
  {Phys. Rev. B}\ }\textbf {\bibinfo {volume} {78}},\ \bibinfo {pages} {184438}
  (\bibinfo {year} {2008})}\BibitemShut {NoStop}%
\bibitem [{\citenamefont {Barth}\ \emph {et~al.}(2010)\citenamefont {Barth},
  \citenamefont {Fecher}, \citenamefont {Balke}, \citenamefont {Ouardi},
  \citenamefont {Graf}, \citenamefont {Felser}, \citenamefont {Shkabko},
  \citenamefont {Weidenkaff}, \citenamefont {Klaer}, \citenamefont {Elmers}
  \emph {et~al.}}]{barth2010itinerant}%
  \BibitemOpen
  \bibfield  {author} {\bibinfo {author} {\bibfnamefont {J.}~\bibnamefont
  {Barth}}, \bibinfo {author} {\bibfnamefont {G.~H.}\ \bibnamefont {Fecher}},
  \bibinfo {author} {\bibfnamefont {B.}~\bibnamefont {Balke}}, \bibinfo
  {author} {\bibfnamefont {S.}~\bibnamefont {Ouardi}}, \bibinfo {author}
  {\bibfnamefont {T.}~\bibnamefont {Graf}}, \bibinfo {author} {\bibfnamefont
  {C.}~\bibnamefont {Felser}}, \bibinfo {author} {\bibfnamefont
  {A.}~\bibnamefont {Shkabko}}, \bibinfo {author} {\bibfnamefont
  {A.}~\bibnamefont {Weidenkaff}}, \bibinfo {author} {\bibfnamefont
  {P.}~\bibnamefont {Klaer}}, \bibinfo {author} {\bibfnamefont {H.~J.}\
  \bibnamefont {Elmers}}, \emph {et~al.},\ }\href@noop {} {\bibfield  {journal}
  {\bibinfo  {journal} {Phys. Rev. B}\ }\textbf {\bibinfo {volume} {81}},\
  \bibinfo {pages} {064404} (\bibinfo {year} {2010})}\BibitemShut {NoStop}%
\bibitem [{\citenamefont {Roy}\ \emph {et~al.}(2019)\citenamefont {Roy},
  \citenamefont {Khan}, \citenamefont {Singha}, \citenamefont {Pariari},\ and\
  \citenamefont {Mandal}}]{roy2019complex}%
  \BibitemOpen
  \bibfield  {author} {\bibinfo {author} {\bibfnamefont {S.}~\bibnamefont
  {Roy}}, \bibinfo {author} {\bibfnamefont {N.}~\bibnamefont {Khan}}, \bibinfo
  {author} {\bibfnamefont {R.}~\bibnamefont {Singha}}, \bibinfo {author}
  {\bibfnamefont {A.}~\bibnamefont {Pariari}},\ and\ \bibinfo {author}
  {\bibfnamefont {P.}~\bibnamefont {Mandal}},\ }\href@noop {} {\bibfield
  {journal} {\bibinfo  {journal} {Phys. Rev. B}\ }\textbf {\bibinfo {volume}
  {99}},\ \bibinfo {pages} {214414} (\bibinfo {year} {2019})}\BibitemShut
  {NoStop}%
\bibitem [{\citenamefont {Felser}\ \emph {et~al.}(2007)\citenamefont {Felser},
  \citenamefont {Fecher},\ and\ \citenamefont {Balke}}]{felser2007spintronics}%
  \BibitemOpen
  \bibfield  {author} {\bibinfo {author} {\bibfnamefont {C.}~\bibnamefont
  {Felser}}, \bibinfo {author} {\bibfnamefont {G.~H.}\ \bibnamefont {Fecher}},\
  and\ \bibinfo {author} {\bibfnamefont {B.}~\bibnamefont {Balke}},\
  }\href@noop {} {\bibfield  {journal} {\bibinfo  {journal} {Angew. Chem. Int.
  Ed.}\ }\textbf {\bibinfo {volume} {46}},\ \bibinfo {pages} {668} (\bibinfo
  {year} {2007})}\BibitemShut {NoStop}%
\bibitem [{\citenamefont {Bombor}\ \emph {et~al.}(2013)\citenamefont {Bombor},
  \citenamefont {Blum}, \citenamefont {Volkonskiy}, \citenamefont {Rodan},
  \citenamefont {Wurmehl}, \citenamefont {Hess},\ and\ \citenamefont
  {B{\"u}chner}}]{bombor2013half}%
  \BibitemOpen
  \bibfield  {author} {\bibinfo {author} {\bibfnamefont {D.}~\bibnamefont
  {Bombor}}, \bibinfo {author} {\bibfnamefont {C.~G.}\ \bibnamefont {Blum}},
  \bibinfo {author} {\bibfnamefont {O.}~\bibnamefont {Volkonskiy}}, \bibinfo
  {author} {\bibfnamefont {S.}~\bibnamefont {Rodan}}, \bibinfo {author}
  {\bibfnamefont {S.}~\bibnamefont {Wurmehl}}, \bibinfo {author} {\bibfnamefont
  {C.}~\bibnamefont {Hess}},\ and\ \bibinfo {author} {\bibfnamefont
  {B.}~\bibnamefont {B{\"u}chner}},\ }\href@noop {} {\bibfield  {journal}
  {\bibinfo  {journal} {Phys. Rev. Lett.}\ }\textbf {\bibinfo {volume} {110}},\
  \bibinfo {pages} {066601} (\bibinfo {year} {2013})}\BibitemShut {NoStop}%
\bibitem [{\citenamefont {Miura}\ \emph {et~al.}(2004)\citenamefont {Miura},
  \citenamefont {Nagao},\ and\ \citenamefont {Shirai}}]{miura2004atomic}%
  \BibitemOpen
  \bibfield  {author} {\bibinfo {author} {\bibfnamefont {Y.}~\bibnamefont
  {Miura}}, \bibinfo {author} {\bibfnamefont {K.}~\bibnamefont {Nagao}},\ and\
  \bibinfo {author} {\bibfnamefont {M.}~\bibnamefont {Shirai}},\ }\href@noop {}
  {\bibfield  {journal} {\bibinfo  {journal} {Phys. Rev. B}\ }\textbf {\bibinfo
  {volume} {69}},\ \bibinfo {pages} {144413} (\bibinfo {year}
  {2004})}\BibitemShut {NoStop}%
\bibitem [{\citenamefont {Kharel}\ \emph {et~al.}(2017)\citenamefont {Kharel},
  \citenamefont {Herran}, \citenamefont {Lukashev}, \citenamefont {Jin},
  \citenamefont {Waybright}, \citenamefont {Gilbert}, \citenamefont {Staten},
  \citenamefont {Gray}, \citenamefont {Valloppilly}, \citenamefont {Huh} \emph
  {et~al.}}]{kharel2017effect}%
  \BibitemOpen
  \bibfield  {author} {\bibinfo {author} {\bibfnamefont {P.}~\bibnamefont
  {Kharel}}, \bibinfo {author} {\bibfnamefont {J.}~\bibnamefont {Herran}},
  \bibinfo {author} {\bibfnamefont {P.}~\bibnamefont {Lukashev}}, \bibinfo
  {author} {\bibfnamefont {Y.}~\bibnamefont {Jin}}, \bibinfo {author}
  {\bibfnamefont {J.}~\bibnamefont {Waybright}}, \bibinfo {author}
  {\bibfnamefont {S.}~\bibnamefont {Gilbert}}, \bibinfo {author} {\bibfnamefont
  {B.}~\bibnamefont {Staten}}, \bibinfo {author} {\bibfnamefont
  {P.}~\bibnamefont {Gray}}, \bibinfo {author} {\bibfnamefont {S.}~\bibnamefont
  {Valloppilly}}, \bibinfo {author} {\bibfnamefont {Y.}~\bibnamefont {Huh}},
  \emph {et~al.},\ }\href@noop {} {\bibfield  {journal} {\bibinfo  {journal}
  {AIP Advances}\ }\textbf {\bibinfo {volume} {7}},\ \bibinfo {pages} {056402}
  (\bibinfo {year} {2017})}\BibitemShut {NoStop}%
\bibitem [{\citenamefont {Mukadam}\ \emph {et~al.}(2016)\citenamefont
  {Mukadam}, \citenamefont {Roy}, \citenamefont {Meena}, \citenamefont
  {Bhatt},\ and\ \citenamefont {Yusuf}}]{mukadam2016quantification}%
  \BibitemOpen
  \bibfield  {author} {\bibinfo {author} {\bibfnamefont {M.}~\bibnamefont
  {Mukadam}}, \bibinfo {author} {\bibfnamefont {S.}~\bibnamefont {Roy}},
  \bibinfo {author} {\bibfnamefont {S.}~\bibnamefont {Meena}}, \bibinfo
  {author} {\bibfnamefont {P.}~\bibnamefont {Bhatt}},\ and\ \bibinfo {author}
  {\bibfnamefont {S.}~\bibnamefont {Yusuf}},\ }\href@noop {} {\bibfield
  {journal} {\bibinfo  {journal} {Phys. Rev. B}\ }\textbf {\bibinfo {volume}
  {94}},\ \bibinfo {pages} {214423} (\bibinfo {year} {2016})}\BibitemShut
  {NoStop}%
\bibitem [{\citenamefont {Bera}\ \emph {et~al.}(2022)\citenamefont {Bera},
  \citenamefont {Mukherjee}, \citenamefont {Mukadam}, \citenamefont {Mondal},
  \citenamefont {Firoz}, \citenamefont {Vaitheeswaran}, \citenamefont {Roy},\
  and\ \citenamefont {Yusuf}}]{bera2022selective}%
  \BibitemOpen
  \bibfield  {author} {\bibinfo {author} {\bibfnamefont {K.}~\bibnamefont
  {Bera}}, \bibinfo {author} {\bibfnamefont {S.}~\bibnamefont {Mukherjee}},
  \bibinfo {author} {\bibfnamefont {M.}~\bibnamefont {Mukadam}}, \bibinfo
  {author} {\bibfnamefont {S.}~\bibnamefont {Mondal}}, \bibinfo {author}
  {\bibfnamefont {M.}~\bibnamefont {Firoz}}, \bibinfo {author} {\bibfnamefont
  {G.}~\bibnamefont {Vaitheeswaran}}, \bibinfo {author} {\bibfnamefont
  {A.}~\bibnamefont {Roy}},\ and\ \bibinfo {author} {\bibfnamefont
  {S.}~\bibnamefont {Yusuf}},\ }\href@noop {} {\bibfield  {journal} {\bibinfo
  {journal} {Applied Physics Letters}\ }\textbf {\bibinfo {volume} {121}},\
  \bibinfo {pages} {052404} (\bibinfo {year} {2022})}\BibitemShut {NoStop}%
\bibitem [{\citenamefont {Bainsla}\ and\ \citenamefont
  {Suresh}(2016)}]{bainsla2016equiatomic}%
  \BibitemOpen
  \bibfield  {author} {\bibinfo {author} {\bibfnamefont {L.}~\bibnamefont
  {Bainsla}}\ and\ \bibinfo {author} {\bibfnamefont {K.}~\bibnamefont
  {Suresh}},\ }\href@noop {} {\bibfield  {journal} {\bibinfo  {journal} {Appl.
  Phys. Rev.}\ }\textbf {\bibinfo {volume} {3}},\ \bibinfo {pages} {031101}
  (\bibinfo {year} {2016})}\BibitemShut {NoStop}%
\bibitem [{\citenamefont {Dai}\ \emph {et~al.}(2009)\citenamefont {Dai},
  \citenamefont {Liu}, \citenamefont {Fecher}, \citenamefont {Felser},
  \citenamefont {Li},\ and\ \citenamefont {Liu}}]{dai2009new}%
  \BibitemOpen
  \bibfield  {author} {\bibinfo {author} {\bibfnamefont {X.}~\bibnamefont
  {Dai}}, \bibinfo {author} {\bibfnamefont {G.}~\bibnamefont {Liu}}, \bibinfo
  {author} {\bibfnamefont {G.~H.}\ \bibnamefont {Fecher}}, \bibinfo {author}
  {\bibfnamefont {C.}~\bibnamefont {Felser}}, \bibinfo {author} {\bibfnamefont
  {Y.}~\bibnamefont {Li}},\ and\ \bibinfo {author} {\bibfnamefont
  {H.}~\bibnamefont {Liu}},\ }\href@noop {} {\bibfield  {journal} {\bibinfo
  {journal} {J. Appl. Phys.}\ }\textbf {\bibinfo {volume} {105}},\ \bibinfo
  {pages} {07E901} (\bibinfo {year} {2009})}\BibitemShut {NoStop}%
\bibitem [{\citenamefont {Alijani}\ \emph
  {et~al.}(2011{\natexlab{a}})\citenamefont {Alijani}, \citenamefont {Ouardi},
  \citenamefont {Fecher}, \citenamefont {Winterlik}, \citenamefont {Naghavi},
  \citenamefont {Kozina}, \citenamefont {Stryganyuk}, \citenamefont {Felser},
  \citenamefont {Ikenaga}, \citenamefont {Yamashita} \emph
  {et~al.}}]{alijani2011electronic}%
  \BibitemOpen
  \bibfield  {author} {\bibinfo {author} {\bibfnamefont {V.}~\bibnamefont
  {Alijani}}, \bibinfo {author} {\bibfnamefont {S.}~\bibnamefont {Ouardi}},
  \bibinfo {author} {\bibfnamefont {G.~H.}\ \bibnamefont {Fecher}}, \bibinfo
  {author} {\bibfnamefont {J.}~\bibnamefont {Winterlik}}, \bibinfo {author}
  {\bibfnamefont {S.~S.}\ \bibnamefont {Naghavi}}, \bibinfo {author}
  {\bibfnamefont {X.}~\bibnamefont {Kozina}}, \bibinfo {author} {\bibfnamefont
  {G.}~\bibnamefont {Stryganyuk}}, \bibinfo {author} {\bibfnamefont
  {C.}~\bibnamefont {Felser}}, \bibinfo {author} {\bibfnamefont
  {E.}~\bibnamefont {Ikenaga}}, \bibinfo {author} {\bibfnamefont
  {Y.}~\bibnamefont {Yamashita}}, \emph {et~al.},\ }\href@noop {} {\bibfield
  {journal} {\bibinfo  {journal} {Phys. Rev. B}\ }\textbf {\bibinfo {volume}
  {84}},\ \bibinfo {pages} {224416} (\bibinfo {year}
  {2011}{\natexlab{a}})}\BibitemShut {NoStop}%
\bibitem [{\citenamefont {Alijani}\ \emph
  {et~al.}(2011{\natexlab{b}})\citenamefont {Alijani}, \citenamefont
  {Winterlik}, \citenamefont {Fecher}, \citenamefont {Naghavi},\ and\
  \citenamefont {Felser}}]{alijani2011quaternary}%
  \BibitemOpen
  \bibfield  {author} {\bibinfo {author} {\bibfnamefont {V.}~\bibnamefont
  {Alijani}}, \bibinfo {author} {\bibfnamefont {J.}~\bibnamefont {Winterlik}},
  \bibinfo {author} {\bibfnamefont {G.~H.}\ \bibnamefont {Fecher}}, \bibinfo
  {author} {\bibfnamefont {S.~S.}\ \bibnamefont {Naghavi}},\ and\ \bibinfo
  {author} {\bibfnamefont {C.}~\bibnamefont {Felser}},\ }\href@noop {}
  {\bibfield  {journal} {\bibinfo  {journal} {Phys. Rev. B}\ }\textbf {\bibinfo
  {volume} {83}},\ \bibinfo {pages} {184428} (\bibinfo {year}
  {2011}{\natexlab{b}})}\BibitemShut {NoStop}%
\bibitem [{\citenamefont {Bainsla}\ \emph
  {et~al.}(2015{\natexlab{a}})\citenamefont {Bainsla}, \citenamefont {Mallick},
  \citenamefont {Raja}, \citenamefont {Nigam}, \citenamefont {Varaprasad},
  \citenamefont {Takahashi}, \citenamefont {Alam}, \citenamefont {Suresh},\
  and\ \citenamefont {Hono}}]{bainsla2015spin}%
  \BibitemOpen
  \bibfield  {author} {\bibinfo {author} {\bibfnamefont {L.}~\bibnamefont
  {Bainsla}}, \bibinfo {author} {\bibfnamefont {A.}~\bibnamefont {Mallick}},
  \bibinfo {author} {\bibfnamefont {M.~M.}\ \bibnamefont {Raja}}, \bibinfo
  {author} {\bibfnamefont {A.}~\bibnamefont {Nigam}}, \bibinfo {author}
  {\bibfnamefont {B.~C.~S.}\ \bibnamefont {Varaprasad}}, \bibinfo {author}
  {\bibfnamefont {Y.}~\bibnamefont {Takahashi}}, \bibinfo {author}
  {\bibfnamefont {A.}~\bibnamefont {Alam}}, \bibinfo {author} {\bibfnamefont
  {K.}~\bibnamefont {Suresh}},\ and\ \bibinfo {author} {\bibfnamefont
  {K.}~\bibnamefont {Hono}},\ }\href@noop {} {\bibfield  {journal} {\bibinfo
  {journal} {Phys. Rev. B}\ }\textbf {\bibinfo {volume} {91}},\ \bibinfo
  {pages} {104408} (\bibinfo {year} {2015}{\natexlab{a}})}\BibitemShut
  {NoStop}%
\bibitem [{\citenamefont {Bainsla}\ \emph
  {et~al.}(2015{\natexlab{b}})\citenamefont {Bainsla}, \citenamefont {Mallick},
  \citenamefont {Raja}, \citenamefont {Coelho}, \citenamefont {Nigam},
  \citenamefont {Johnson}, \citenamefont {Alam},\ and\ \citenamefont
  {Suresh}}]{bainsla2015origin}%
  \BibitemOpen
  \bibfield  {author} {\bibinfo {author} {\bibfnamefont {L.}~\bibnamefont
  {Bainsla}}, \bibinfo {author} {\bibfnamefont {A.}~\bibnamefont {Mallick}},
  \bibinfo {author} {\bibfnamefont {M.~M.}\ \bibnamefont {Raja}}, \bibinfo
  {author} {\bibfnamefont {A.}~\bibnamefont {Coelho}}, \bibinfo {author}
  {\bibfnamefont {A.}~\bibnamefont {Nigam}}, \bibinfo {author} {\bibfnamefont
  {D.~D.}\ \bibnamefont {Johnson}}, \bibinfo {author} {\bibfnamefont
  {A.}~\bibnamefont {Alam}},\ and\ \bibinfo {author} {\bibfnamefont
  {K.}~\bibnamefont {Suresh}},\ }\href@noop {} {\bibfield  {journal} {\bibinfo
  {journal} {Phys. Rev. B}\ }\textbf {\bibinfo {volume} {92}},\ \bibinfo
  {pages} {045201} (\bibinfo {year} {2015}{\natexlab{b}})}\BibitemShut
  {NoStop}%
\bibitem [{\citenamefont {Gupta}\ \emph {et~al.}(2022)\citenamefont {Gupta},
  \citenamefont {Chakraborty}, \citenamefont {Pakhira}, \citenamefont
  {Barreteau}, \citenamefont {Crivello}, \citenamefont {Bandyopadhyay},
  \citenamefont {Greneche}, \citenamefont {Alleno},\ and\ \citenamefont
  {Mazumdar}}]{gupta2022coexisting}%
  \BibitemOpen
  \bibfield  {author} {\bibinfo {author} {\bibfnamefont {S.}~\bibnamefont
  {Gupta}}, \bibinfo {author} {\bibfnamefont {S.}~\bibnamefont {Chakraborty}},
  \bibinfo {author} {\bibfnamefont {S.}~\bibnamefont {Pakhira}}, \bibinfo
  {author} {\bibfnamefont {C.}~\bibnamefont {Barreteau}}, \bibinfo {author}
  {\bibfnamefont {J.-C.}\ \bibnamefont {Crivello}}, \bibinfo {author}
  {\bibfnamefont {B.}~\bibnamefont {Bandyopadhyay}}, \bibinfo {author}
  {\bibfnamefont {J.~M.}\ \bibnamefont {Greneche}}, \bibinfo {author}
  {\bibfnamefont {E.}~\bibnamefont {Alleno}},\ and\ \bibinfo {author}
  {\bibfnamefont {C.}~\bibnamefont {Mazumdar}},\ }\href@noop {} {\bibfield
  {journal} {\bibinfo  {journal} {Physical Review B}\ }\textbf {\bibinfo
  {volume} {106}},\ \bibinfo {pages} {115148} (\bibinfo {year}
  {2022})}\BibitemShut {NoStop}%
\bibitem [{\citenamefont {Gupta}\ \emph
  {et~al.}(2023{\natexlab{a}})\citenamefont {Gupta}, \citenamefont
  {Chakraborty}, \citenamefont {Bhasin}, \citenamefont {Pakhira}, \citenamefont
  {Dan}, \citenamefont {Barreteau}, \citenamefont {Crivello}, \citenamefont
  {Jha}, \citenamefont {Avdeev}, \citenamefont {Greneche} \emph
  {et~al.}}]{gupta2023high}%
  \BibitemOpen
  \bibfield  {author} {\bibinfo {author} {\bibfnamefont {S.}~\bibnamefont
  {Gupta}}, \bibinfo {author} {\bibfnamefont {S.}~\bibnamefont {Chakraborty}},
  \bibinfo {author} {\bibfnamefont {V.}~\bibnamefont {Bhasin}}, \bibinfo
  {author} {\bibfnamefont {S.}~\bibnamefont {Pakhira}}, \bibinfo {author}
  {\bibfnamefont {S.}~\bibnamefont {Dan}}, \bibinfo {author} {\bibfnamefont
  {C.}~\bibnamefont {Barreteau}}, \bibinfo {author} {\bibfnamefont {J.-C.}\
  \bibnamefont {Crivello}}, \bibinfo {author} {\bibfnamefont {S.}~\bibnamefont
  {Jha}}, \bibinfo {author} {\bibfnamefont {M.}~\bibnamefont {Avdeev}},
  \bibinfo {author} {\bibfnamefont {J.~M.}\ \bibnamefont {Greneche}}, \emph
  {et~al.},\ }\href@noop {} {\bibfield  {journal} {\bibinfo  {journal} {arXiv
  preprint arXiv:2303.08579}\ } (\bibinfo {year}
  {2023}{\natexlab{a}})}\BibitemShut {NoStop}%
\bibitem [{\citenamefont {Gupta}\ \emph
  {et~al.}(2023{\natexlab{b}})\citenamefont {Gupta}, \citenamefont {Sau},
  \citenamefont {Kumar},\ and\ \citenamefont {Mazumdar}}]{gupta2023rare}%
  \BibitemOpen
  \bibfield  {author} {\bibinfo {author} {\bibfnamefont {S.}~\bibnamefont
  {Gupta}}, \bibinfo {author} {\bibfnamefont {J.}~\bibnamefont {Sau}}, \bibinfo
  {author} {\bibfnamefont {M.}~\bibnamefont {Kumar}},\ and\ \bibinfo {author}
  {\bibfnamefont {C.}~\bibnamefont {Mazumdar}},\ }\href@noop {} {\bibfield
  {journal} {\bibinfo  {journal} {arXiv preprint arXiv:2303.08589}\ } (\bibinfo
  {year} {2023}{\natexlab{b}})}\BibitemShut {NoStop}%
\bibitem [{\citenamefont {Rani}\ \emph {et~al.}(2017)\citenamefont {Rani},
  \citenamefont {Suresh}, \citenamefont {Yadav}, \citenamefont {Jha},
  \citenamefont {Bhattacharyya}, \citenamefont {Varma},\ and\ \citenamefont
  {Alam}}]{rani2017structural}%
  \BibitemOpen
  \bibfield  {author} {\bibinfo {author} {\bibfnamefont {D.}~\bibnamefont
  {Rani}}, \bibinfo {author} {\bibfnamefont {K.~G.}\ \bibnamefont {Suresh}},
  \bibinfo {author} {\bibfnamefont {A.~K.}\ \bibnamefont {Yadav}}, \bibinfo
  {author} {\bibfnamefont {S.~N.}\ \bibnamefont {Jha}}, \bibinfo {author}
  {\bibfnamefont {D.}~\bibnamefont {Bhattacharyya}}, \bibinfo {author}
  {\bibfnamefont {M.~R.}\ \bibnamefont {Varma}},\ and\ \bibinfo {author}
  {\bibfnamefont {A.}~\bibnamefont {Alam}},\ }\href@noop {} {\bibfield
  {journal} {\bibinfo  {journal} {Phys. Rev. B}\ }\textbf {\bibinfo {volume}
  {96}},\ \bibinfo {pages} {184404} (\bibinfo {year} {2017})}\BibitemShut
  {NoStop}%
\bibitem [{\citenamefont {Bainsla}\ \emph
  {et~al.}(2015{\natexlab{c}})\citenamefont {Bainsla}, \citenamefont {Raja},
  \citenamefont {Nigam},\ and\ \citenamefont {Suresh}}]{bainsla2015corufex}%
  \BibitemOpen
  \bibfield  {author} {\bibinfo {author} {\bibfnamefont {L.}~\bibnamefont
  {Bainsla}}, \bibinfo {author} {\bibfnamefont {M.~M.}\ \bibnamefont {Raja}},
  \bibinfo {author} {\bibfnamefont {A.~K.}\ \bibnamefont {Nigam}},\ and\
  \bibinfo {author} {\bibfnamefont {K.~G.}\ \bibnamefont {Suresh}},\
  }\href@noop {} {\bibfield  {journal} {\bibinfo  {journal} {J. Alloys Compd.}\
  }\textbf {\bibinfo {volume} {651}},\ \bibinfo {pages} {631} (\bibinfo {year}
  {2015}{\natexlab{c}})}\BibitemShut {NoStop}%
\bibitem [{\citenamefont {Bainsla}\ \emph {et~al.}(2014)\citenamefont
  {Bainsla}, \citenamefont {Suresh}, \citenamefont {Nigam}, \citenamefont
  {Manivel~Raja}, \citenamefont {Varaprasad}, \citenamefont {Takahashi},\ and\
  \citenamefont {Hono}}]{bainsla2014high}%
  \BibitemOpen
  \bibfield  {author} {\bibinfo {author} {\bibfnamefont {L.}~\bibnamefont
  {Bainsla}}, \bibinfo {author} {\bibfnamefont {K.}~\bibnamefont {Suresh}},
  \bibinfo {author} {\bibfnamefont {A.}~\bibnamefont {Nigam}}, \bibinfo
  {author} {\bibfnamefont {M.}~\bibnamefont {Manivel~Raja}}, \bibinfo {author}
  {\bibfnamefont {B.~C.~S.}\ \bibnamefont {Varaprasad}}, \bibinfo {author}
  {\bibfnamefont {Y.}~\bibnamefont {Takahashi}},\ and\ \bibinfo {author}
  {\bibfnamefont {K.}~\bibnamefont {Hono}},\ }\href@noop {} {\bibfield
  {journal} {\bibinfo  {journal} {J. Appl. Phys.}\ }\textbf {\bibinfo {volume}
  {116}},\ \bibinfo {pages} {203902} (\bibinfo {year} {2014})}\BibitemShut
  {NoStop}%
\bibitem [{\citenamefont
  {Rodr{\'\i}guez-Carvajal}(1993)}]{rodriguez1993recent}%
  \BibitemOpen
  \bibfield  {author} {\bibinfo {author} {\bibfnamefont {J.}~\bibnamefont
  {Rodr{\'\i}guez-Carvajal}},\ }\href@noop {} {\bibfield  {journal} {\bibinfo
  {journal} {Phys. B: Condens. Matter}\ }\textbf {\bibinfo {volume} {192}},\
  \bibinfo {pages} {55} (\bibinfo {year} {1993})}\BibitemShut {NoStop}%
\bibitem [{\citenamefont {Bl{\"o}chl}(1994)}]{blochl1994projector}%
  \BibitemOpen
  \bibfield  {author} {\bibinfo {author} {\bibfnamefont {P.~E.}\ \bibnamefont
  {Bl{\"o}chl}},\ }\href@noop {} {\bibfield  {journal} {\bibinfo  {journal}
  {Phys. Rev. B}\ }\textbf {\bibinfo {volume} {50}},\ \bibinfo {pages} {17953}
  (\bibinfo {year} {1994})}\BibitemShut {NoStop}%
\bibitem [{\citenamefont {Kresse}\ and\ \citenamefont
  {Hafner}(1993)}]{kresse1993ab}%
  \BibitemOpen
  \bibfield  {author} {\bibinfo {author} {\bibfnamefont {G.}~\bibnamefont
  {Kresse}}\ and\ \bibinfo {author} {\bibfnamefont {J.}~\bibnamefont
  {Hafner}},\ }\href@noop {} {\bibfield  {journal} {\bibinfo  {journal} {Phys.
  Rev. B}\ }\textbf {\bibinfo {volume} {48}},\ \bibinfo {pages} {13115}
  (\bibinfo {year} {1993})}\BibitemShut {NoStop}%
\bibitem [{\citenamefont {Kresse}\ and\ \citenamefont
  {Hafner}(1994)}]{kresse1994norm}%
  \BibitemOpen
  \bibfield  {author} {\bibinfo {author} {\bibfnamefont {G.}~\bibnamefont
  {Kresse}}\ and\ \bibinfo {author} {\bibfnamefont {J.}~\bibnamefont
  {Hafner}},\ }\href@noop {} {\bibfield  {journal} {\bibinfo  {journal} {J.
  Condens. Matter Phys.}\ }\textbf {\bibinfo {volume} {6}},\ \bibinfo {pages}
  {8245} (\bibinfo {year} {1994})}\BibitemShut {NoStop}%
\bibitem [{\citenamefont {Perdew}\ \emph {et~al.}(1996)\citenamefont {Perdew},
  \citenamefont {Burke},\ and\ \citenamefont
  {Ernzerhof}}]{perdew1996generalized}%
  \BibitemOpen
  \bibfield  {author} {\bibinfo {author} {\bibfnamefont {J.~P.}\ \bibnamefont
  {Perdew}}, \bibinfo {author} {\bibfnamefont {K.}~\bibnamefont {Burke}},\ and\
  \bibinfo {author} {\bibfnamefont {M.}~\bibnamefont {Ernzerhof}},\ }\href@noop
  {} {\bibfield  {journal} {\bibinfo  {journal} {Phys. Rev. Lett.}\ }\textbf
  {\bibinfo {volume} {77}},\ \bibinfo {pages} {3865} (\bibinfo {year}
  {1996})}\BibitemShut {NoStop}%
\bibitem [{\citenamefont {Bl{\"o}chl}\ \emph {et~al.}(1994)\citenamefont
  {Bl{\"o}chl}, \citenamefont {Jepsen},\ and\ \citenamefont
  {Andersen}}]{blochl1994improved}%
  \BibitemOpen
  \bibfield  {author} {\bibinfo {author} {\bibfnamefont {P.~E.}\ \bibnamefont
  {Bl{\"o}chl}}, \bibinfo {author} {\bibfnamefont {O.}~\bibnamefont {Jepsen}},\
  and\ \bibinfo {author} {\bibfnamefont {O.~K.}\ \bibnamefont {Andersen}},\
  }\href@noop {} {\bibfield  {journal} {\bibinfo  {journal} {Phys. Rev. B}\
  }\textbf {\bibinfo {volume} {49}},\ \bibinfo {pages} {16223} (\bibinfo {year}
  {1994})}\BibitemShut {NoStop}%
\bibitem [{\citenamefont {Zunger}\ \emph {et~al.}(1990)\citenamefont {Zunger},
  \citenamefont {Wei}, \citenamefont {Ferreira},\ and\ \citenamefont
  {Bernard}}]{zunger1990special}%
  \BibitemOpen
  \bibfield  {author} {\bibinfo {author} {\bibfnamefont {A.}~\bibnamefont
  {Zunger}}, \bibinfo {author} {\bibfnamefont {S.-H.}\ \bibnamefont {Wei}},
  \bibinfo {author} {\bibfnamefont {L.}~\bibnamefont {Ferreira}},\ and\
  \bibinfo {author} {\bibfnamefont {J.~E.}\ \bibnamefont {Bernard}},\
  }\href@noop {} {\bibfield  {journal} {\bibinfo  {journal} {Phys. Rev. Lett.}\
  }\textbf {\bibinfo {volume} {65}},\ \bibinfo {pages} {353} (\bibinfo {year}
  {1990})}\BibitemShut {NoStop}%
\bibitem [{\citenamefont {Sanchez}\ \emph {et~al.}(1984)\citenamefont
  {Sanchez}, \citenamefont {Ducastelle},\ and\ \citenamefont
  {Gratias}}]{sanchez1984generalized}%
  \BibitemOpen
  \bibfield  {author} {\bibinfo {author} {\bibfnamefont {J.~M.}\ \bibnamefont
  {Sanchez}}, \bibinfo {author} {\bibfnamefont {F.}~\bibnamefont
  {Ducastelle}},\ and\ \bibinfo {author} {\bibfnamefont {D.}~\bibnamefont
  {Gratias}},\ }\href@noop {} {\bibfield  {journal} {\bibinfo  {journal} {Phys.
  A: Stat. Mech. Appl.}\ }\textbf {\bibinfo {volume} {128}},\ \bibinfo {pages}
  {334} (\bibinfo {year} {1984})}\BibitemShut {NoStop}%
\bibitem [{\citenamefont {Van De~Walle}(2009)}]{van2009multicomponent}%
  \BibitemOpen
  \bibfield  {author} {\bibinfo {author} {\bibfnamefont {A.}~\bibnamefont {Van
  De~Walle}},\ }\href@noop {} {\bibfield  {journal} {\bibinfo  {journal}
  {Calphad}\ }\textbf {\bibinfo {volume} {33}},\ \bibinfo {pages} {266}
  (\bibinfo {year} {2009})}\BibitemShut {NoStop}%
\bibitem [{\citenamefont {Van~de Walle}\ \emph {et~al.}(2013)\citenamefont
  {Van~de Walle}, \citenamefont {Tiwary}, \citenamefont {De~Jong},
  \citenamefont {Olmsted}, \citenamefont {Asta}, \citenamefont {Dick},
  \citenamefont {Shin}, \citenamefont {Wang}, \citenamefont {Chen},\ and\
  \citenamefont {Liu}}]{van2013efficient}%
  \BibitemOpen
  \bibfield  {author} {\bibinfo {author} {\bibfnamefont {A.}~\bibnamefont
  {Van~de Walle}}, \bibinfo {author} {\bibfnamefont {P.}~\bibnamefont
  {Tiwary}}, \bibinfo {author} {\bibfnamefont {M.}~\bibnamefont {De~Jong}},
  \bibinfo {author} {\bibfnamefont {D.}~\bibnamefont {Olmsted}}, \bibinfo
  {author} {\bibfnamefont {M.}~\bibnamefont {Asta}}, \bibinfo {author}
  {\bibfnamefont {A.}~\bibnamefont {Dick}}, \bibinfo {author} {\bibfnamefont
  {D.}~\bibnamefont {Shin}}, \bibinfo {author} {\bibfnamefont {Y.}~\bibnamefont
  {Wang}}, \bibinfo {author} {\bibfnamefont {L.-Q.}\ \bibnamefont {Chen}},\
  and\ \bibinfo {author} {\bibfnamefont {Z.-K.}\ \bibnamefont {Liu}},\
  }\href@noop {} {\bibfield  {journal} {\bibinfo  {journal} {Calphad}\ }\textbf
  {\bibinfo {volume} {42}},\ \bibinfo {pages} {13} (\bibinfo {year}
  {2013})}\BibitemShut {NoStop}%
\bibitem [{\citenamefont {Galanakis}\ \emph {et~al.}(2002)\citenamefont
  {Galanakis}, \citenamefont {Dederichs},\ and\ \citenamefont
  {Papanikolaou}}]{galanakis2002slater}%
  \BibitemOpen
  \bibfield  {author} {\bibinfo {author} {\bibfnamefont {I.}~\bibnamefont
  {Galanakis}}, \bibinfo {author} {\bibfnamefont {P.}~\bibnamefont
  {Dederichs}},\ and\ \bibinfo {author} {\bibfnamefont {N.}~\bibnamefont
  {Papanikolaou}},\ }\href@noop {} {\bibfield  {journal} {\bibinfo  {journal}
  {Phys. Rev. B}\ }\textbf {\bibinfo {volume} {66}},\ \bibinfo {pages} {174429}
  (\bibinfo {year} {2002})}\BibitemShut {NoStop}%
\bibitem [{\citenamefont {{\"O}zdo{\u{g}}an}\ \emph {et~al.}(2013)\citenamefont
  {{\"O}zdo{\u{g}}an}, \citenamefont {{\c{S}}a{\c{s}}{\i}o{\u{g}}lu},\ and\
  \citenamefont {Galanakis}}]{ozdougan2013slater}%
  \BibitemOpen
  \bibfield  {author} {\bibinfo {author} {\bibfnamefont {K.}~\bibnamefont
  {{\"O}zdo{\u{g}}an}}, \bibinfo {author} {\bibfnamefont {E.}~\bibnamefont
  {{\c{S}}a{\c{s}}{\i}o{\u{g}}lu}},\ and\ \bibinfo {author} {\bibfnamefont
  {I.}~\bibnamefont {Galanakis}},\ }\href@noop {} {\bibfield  {journal}
  {\bibinfo  {journal} {J. Appl. Phys.}\ }\textbf {\bibinfo {volume} {113}},\
  \bibinfo {pages} {193903} (\bibinfo {year} {2013})}\BibitemShut {NoStop}%
\bibitem [{\citenamefont {Webster}\ and\ \citenamefont
  {Ziebeck}(1973)}]{webster1973magnetic}%
  \BibitemOpen
  \bibfield  {author} {\bibinfo {author} {\bibfnamefont {P.}~\bibnamefont
  {Webster}}\ and\ \bibinfo {author} {\bibfnamefont {K.}~\bibnamefont
  {Ziebeck}},\ }\href@noop {} {\bibfield  {journal} {\bibinfo  {journal} {J
  Phys Chem Solids}\ }\textbf {\bibinfo {volume} {34}},\ \bibinfo {pages}
  {1647} (\bibinfo {year} {1973})}\BibitemShut {NoStop}%
\bibitem [{\citenamefont {Venkateswara}\ \emph {et~al.}(2020)\citenamefont
  {Venkateswara}, \citenamefont {Rani}, \citenamefont {Suresh},\ and\
  \citenamefont {Alam}}]{venkateswara2020half}%
  \BibitemOpen
  \bibfield  {author} {\bibinfo {author} {\bibfnamefont {Y.}~\bibnamefont
  {Venkateswara}}, \bibinfo {author} {\bibfnamefont {D.}~\bibnamefont {Rani}},
  \bibinfo {author} {\bibfnamefont {K.}~\bibnamefont {Suresh}},\ and\ \bibinfo
  {author} {\bibfnamefont {A.}~\bibnamefont {Alam}},\ }\href@noop {} {\bibfield
   {journal} {\bibinfo  {journal} {J. Magn. Magn}\ }\textbf {\bibinfo {volume}
  {502}},\ \bibinfo {pages} {166536} (\bibinfo {year} {2020})}\BibitemShut
  {NoStop}%
\bibitem [{\citenamefont {Balke}\ \emph {et~al.}(2007)\citenamefont {Balke},
  \citenamefont {Wurmehl}, \citenamefont {Fecher}, \citenamefont {Felser},
  \citenamefont {Alves}, \citenamefont {Bernardi},\ and\ \citenamefont
  {Morais}}]{balke2007structural}%
  \BibitemOpen
  \bibfield  {author} {\bibinfo {author} {\bibfnamefont {B.}~\bibnamefont
  {Balke}}, \bibinfo {author} {\bibfnamefont {S.}~\bibnamefont {Wurmehl}},
  \bibinfo {author} {\bibfnamefont {G.~H.}\ \bibnamefont {Fecher}}, \bibinfo
  {author} {\bibfnamefont {C.}~\bibnamefont {Felser}}, \bibinfo {author}
  {\bibfnamefont {M.~C.}\ \bibnamefont {Alves}}, \bibinfo {author}
  {\bibfnamefont {F.}~\bibnamefont {Bernardi}},\ and\ \bibinfo {author}
  {\bibfnamefont {J.}~\bibnamefont {Morais}},\ }\href@noop {} {\bibfield
  {journal} {\bibinfo  {journal} {Appl. Phys. Lett.}\ }\textbf {\bibinfo
  {volume} {90}},\ \bibinfo {pages} {172501} (\bibinfo {year}
  {2007})}\BibitemShut {NoStop}%
\bibitem [{\citenamefont {Bainsla}\ \emph
  {et~al.}(2015{\natexlab{d}})\citenamefont {Bainsla}, \citenamefont {Yadav},
  \citenamefont {Venkateswara}, \citenamefont {Jha}, \citenamefont
  {Bhattacharyya},\ and\ \citenamefont {Suresh}}]{bainsla2015local}%
  \BibitemOpen
  \bibfield  {author} {\bibinfo {author} {\bibfnamefont {L.}~\bibnamefont
  {Bainsla}}, \bibinfo {author} {\bibfnamefont {A.}~\bibnamefont {Yadav}},
  \bibinfo {author} {\bibfnamefont {Y.}~\bibnamefont {Venkateswara}}, \bibinfo
  {author} {\bibfnamefont {S.}~\bibnamefont {Jha}}, \bibinfo {author}
  {\bibfnamefont {D.}~\bibnamefont {Bhattacharyya}},\ and\ \bibinfo {author}
  {\bibfnamefont {K.}~\bibnamefont {Suresh}},\ }\href@noop {} {\bibfield
  {journal} {\bibinfo  {journal} {J. Alloys Compd.}\ }\textbf {\bibinfo
  {volume} {651}},\ \bibinfo {pages} {509} (\bibinfo {year}
  {2015}{\natexlab{d}})}\BibitemShut {NoStop}%
\bibitem [{\citenamefont {Ravel}\ \emph {et~al.}(2002)\citenamefont {Ravel},
  \citenamefont {Raphael}, \citenamefont {Harris},\ and\ \citenamefont
  {Huang}}]{ravel2002exafs}%
  \BibitemOpen
  \bibfield  {author} {\bibinfo {author} {\bibfnamefont {B.}~\bibnamefont
  {Ravel}}, \bibinfo {author} {\bibfnamefont {M.}~\bibnamefont {Raphael}},
  \bibinfo {author} {\bibfnamefont {V.}~\bibnamefont {Harris}},\ and\ \bibinfo
  {author} {\bibfnamefont {Q.}~\bibnamefont {Huang}},\ }\href@noop {}
  {\bibfield  {journal} {\bibinfo  {journal} {Phys. Rev. B}\ }\textbf {\bibinfo
  {volume} {65}},\ \bibinfo {pages} {184431} (\bibinfo {year}
  {2002})}\BibitemShut {NoStop}%
\bibitem [{\citenamefont {Basu}\ \emph {et~al.}(2014)\citenamefont {Basu},
  \citenamefont {Nayak}, \citenamefont {Yadav}, \citenamefont {Agrawal},
  \citenamefont {Poswal}, \citenamefont {Bhattacharyya}, \citenamefont {Jha},\
  and\ \citenamefont {Sahoo}}]{basu2014comprehensive}%
  \BibitemOpen
  \bibfield  {author} {\bibinfo {author} {\bibfnamefont {S.}~\bibnamefont
  {Basu}}, \bibinfo {author} {\bibfnamefont {C.}~\bibnamefont {Nayak}},
  \bibinfo {author} {\bibfnamefont {A.}~\bibnamefont {Yadav}}, \bibinfo
  {author} {\bibfnamefont {A.}~\bibnamefont {Agrawal}}, \bibinfo {author}
  {\bibfnamefont {A.}~\bibnamefont {Poswal}}, \bibinfo {author} {\bibfnamefont
  {D.}~\bibnamefont {Bhattacharyya}}, \bibinfo {author} {\bibfnamefont
  {S.}~\bibnamefont {Jha}},\ and\ \bibinfo {author} {\bibfnamefont
  {N.}~\bibnamefont {Sahoo}},\ }in\ \href@noop {} {\emph {\bibinfo {booktitle}
  {J. Phys. Conf. Ser.}}},\ Vol.\ \bibinfo {volume} {493}\ (\bibinfo {year}
  {2014})\ p.\ \bibinfo {pages} {012032}\BibitemShut {NoStop}%
\bibitem [{\citenamefont {Poswal}\ \emph {et~al.}(2014)\citenamefont {Poswal},
  \citenamefont {Agrawal}, \citenamefont {Yadav}, \citenamefont {Nayak},
  \citenamefont {Basu}, \citenamefont {Kane}, \citenamefont {Garg},
  \citenamefont {Bhattachryya}, \citenamefont {Jha},\ and\ \citenamefont
  {Sahoo}}]{poswal2014commissioning}%
  \BibitemOpen
  \bibfield  {author} {\bibinfo {author} {\bibfnamefont {A.}~\bibnamefont
  {Poswal}}, \bibinfo {author} {\bibfnamefont {A.}~\bibnamefont {Agrawal}},
  \bibinfo {author} {\bibfnamefont {A.}~\bibnamefont {Yadav}}, \bibinfo
  {author} {\bibfnamefont {C.}~\bibnamefont {Nayak}}, \bibinfo {author}
  {\bibfnamefont {S.}~\bibnamefont {Basu}}, \bibinfo {author} {\bibfnamefont
  {S.}~\bibnamefont {Kane}}, \bibinfo {author} {\bibfnamefont {C.}~\bibnamefont
  {Garg}}, \bibinfo {author} {\bibfnamefont {D.}~\bibnamefont {Bhattachryya}},
  \bibinfo {author} {\bibfnamefont {S.}~\bibnamefont {Jha}},\ and\ \bibinfo
  {author} {\bibfnamefont {N.}~\bibnamefont {Sahoo}},\ }in\ \href@noop {}
  {\emph {\bibinfo {booktitle} {AIP Conf Proc .}}},\ Vol.\ \bibinfo {volume}
  {1591}\ (\bibinfo {year} {2014})\ pp.\ \bibinfo {pages}
  {649--651}\BibitemShut {NoStop}%
\bibitem [{\citenamefont {Konigsberger}\ and\ \citenamefont
  {Prins}(1988)}]{konigsberger1988x}%
  \BibitemOpen
  \bibfield  {author} {\bibinfo {author} {\bibfnamefont {D.}~\bibnamefont
  {Konigsberger}}\ and\ \bibinfo {author} {\bibfnamefont {R.}~\bibnamefont
  {Prins}},\ }\href@noop {} {\bibfield  {journal} {\bibinfo  {journal}
  {Wiley/Interscience, New York}\ }\textbf {\bibinfo {volume} {159}},\ \bibinfo
  {pages} {160} (\bibinfo {year} {1988})}\BibitemShut {NoStop}%
\bibitem [{\citenamefont {Ravel}\ and\ \citenamefont
  {Newville}(2005)}]{ravel2005athena}%
  \BibitemOpen
  \bibfield  {author} {\bibinfo {author} {\bibfnamefont {B.}~\bibnamefont
  {Ravel}}\ and\ \bibinfo {author} {\bibfnamefont {M.}~\bibnamefont
  {Newville}},\ }\href@noop {} {\bibfield  {journal} {\bibinfo  {journal} {J.
  Synchrotron Radiat.}\ }\textbf {\bibinfo {volume} {12}},\ \bibinfo {pages}
  {537} (\bibinfo {year} {2005})}\BibitemShut {NoStop}%
\bibitem [{\citenamefont {Samanta}\ \emph {et~al.}(2018)\citenamefont
  {Samanta}, \citenamefont {Bhobe}, \citenamefont {Das}, \citenamefont
  {Kumar},\ and\ \citenamefont {Nigam}}]{samanta2018reentrant}%
  \BibitemOpen
  \bibfield  {author} {\bibinfo {author} {\bibfnamefont {T.}~\bibnamefont
  {Samanta}}, \bibinfo {author} {\bibfnamefont {P.}~\bibnamefont {Bhobe}},
  \bibinfo {author} {\bibfnamefont {A.}~\bibnamefont {Das}}, \bibinfo {author}
  {\bibfnamefont {A.}~\bibnamefont {Kumar}},\ and\ \bibinfo {author}
  {\bibfnamefont {A.}~\bibnamefont {Nigam}},\ }\href@noop {} {\bibfield
  {journal} {\bibinfo  {journal} {Phys. Rev. B}\ }\textbf {\bibinfo {volume}
  {97}},\ \bibinfo {pages} {184421} (\bibinfo {year} {2018})}\BibitemShut
  {NoStop}%
\bibitem [{\citenamefont {Kroder}\ \emph {et~al.}(2019)\citenamefont {Kroder},
  \citenamefont {Manna}, \citenamefont {Kriegner}, \citenamefont {Sukhanov},
  \citenamefont {Liu}, \citenamefont {Borrmann}, \citenamefont {Hoser},
  \citenamefont {Gooth}, \citenamefont {Schnelle}, \citenamefont {Inosov} \emph
  {et~al.}}]{kroder2019spin}%
  \BibitemOpen
  \bibfield  {author} {\bibinfo {author} {\bibfnamefont {J.}~\bibnamefont
  {Kroder}}, \bibinfo {author} {\bibfnamefont {K.}~\bibnamefont {Manna}},
  \bibinfo {author} {\bibfnamefont {D.}~\bibnamefont {Kriegner}}, \bibinfo
  {author} {\bibfnamefont {A.}~\bibnamefont {Sukhanov}}, \bibinfo {author}
  {\bibfnamefont {E.}~\bibnamefont {Liu}}, \bibinfo {author} {\bibfnamefont
  {H.}~\bibnamefont {Borrmann}}, \bibinfo {author} {\bibfnamefont
  {A.}~\bibnamefont {Hoser}}, \bibinfo {author} {\bibfnamefont
  {J.}~\bibnamefont {Gooth}}, \bibinfo {author} {\bibfnamefont
  {W.}~\bibnamefont {Schnelle}}, \bibinfo {author} {\bibfnamefont {D.~S.}\
  \bibnamefont {Inosov}}, \emph {et~al.},\ }\href@noop {} {\bibfield  {journal}
  {\bibinfo  {journal} {Phys. Rev. B}\ }\textbf {\bibinfo {volume} {99}},\
  \bibinfo {pages} {174410} (\bibinfo {year} {2019})}\BibitemShut {NoStop}%
\bibitem [{\citenamefont {Zhang}\ \emph {et~al.}(2014)\citenamefont {Zhang},
  \citenamefont {Sun}, \citenamefont {Wang}, \citenamefont {Li}, \citenamefont
  {Zhang}, \citenamefont {Sui}, \citenamefont {Luo}, \citenamefont {Meng},
  \citenamefont {Qian},\ and\ \citenamefont {Wu}}]{zhang2014spin}%
  \BibitemOpen
  \bibfield  {author} {\bibinfo {author} {\bibfnamefont {W.}~\bibnamefont
  {Zhang}}, \bibinfo {author} {\bibfnamefont {Y.}~\bibnamefont {Sun}}, \bibinfo
  {author} {\bibfnamefont {H.}~\bibnamefont {Wang}}, \bibinfo {author}
  {\bibfnamefont {Y.}~\bibnamefont {Li}}, \bibinfo {author} {\bibfnamefont
  {X.}~\bibnamefont {Zhang}}, \bibinfo {author} {\bibfnamefont
  {Y.}~\bibnamefont {Sui}}, \bibinfo {author} {\bibfnamefont {H.}~\bibnamefont
  {Luo}}, \bibinfo {author} {\bibfnamefont {F.}~\bibnamefont {Meng}}, \bibinfo
  {author} {\bibfnamefont {Z.}~\bibnamefont {Qian}},\ and\ \bibinfo {author}
  {\bibfnamefont {G.}~\bibnamefont {Wu}},\ }\href@noop {} {\bibfield  {journal}
  {\bibinfo  {journal} {J. Alloys Compd.}\ }\textbf {\bibinfo {volume} {589}},\
  \bibinfo {pages} {230} (\bibinfo {year} {2014})}\BibitemShut {NoStop}%
\bibitem [{\citenamefont {Das}\ \emph {et~al.}(2003)\citenamefont {Das},
  \citenamefont {Paranjpe},\ and\ \citenamefont {Murayama}}]{das2003neutron}%
  \BibitemOpen
  \bibfield  {author} {\bibinfo {author} {\bibfnamefont {A.}~\bibnamefont
  {Das}}, \bibinfo {author} {\bibfnamefont {S.}~\bibnamefont {Paranjpe}},\ and\
  \bibinfo {author} {\bibfnamefont {S.}~\bibnamefont {Murayama}},\ }\href@noop
  {} {\bibfield  {journal} {\bibinfo  {journal} {Phys. B: Condens. Matter}\
  }\textbf {\bibinfo {volume} {335}},\ \bibinfo {pages} {130} (\bibinfo {year}
  {2003})}\BibitemShut {NoStop}%
\bibitem [{\citenamefont {Das}\ \emph {et~al.}(1999)\citenamefont {Das},
  \citenamefont {Paranjpe}, \citenamefont {Honda}, \citenamefont {Murayama},\
  and\ \citenamefont {Tsuchiya}}]{das1999neutron}%
  \BibitemOpen
  \bibfield  {author} {\bibinfo {author} {\bibfnamefont {A.}~\bibnamefont
  {Das}}, \bibinfo {author} {\bibfnamefont {S.}~\bibnamefont {Paranjpe}},
  \bibinfo {author} {\bibfnamefont {S.}~\bibnamefont {Honda}}, \bibinfo
  {author} {\bibfnamefont {S.}~\bibnamefont {Murayama}},\ and\ \bibinfo
  {author} {\bibfnamefont {Y.}~\bibnamefont {Tsuchiya}},\ }\href@noop {}
  {\bibfield  {journal} {\bibinfo  {journal} {J. Condens. Matter Phys.}\
  }\textbf {\bibinfo {volume} {11}},\ \bibinfo {pages} {5209} (\bibinfo {year}
  {1999})}\BibitemShut {NoStop}%
\bibitem [{\citenamefont {Halpern}\ and\ \citenamefont
  {Holstein}(1941)}]{halpern1941passage}%
  \BibitemOpen
  \bibfield  {author} {\bibinfo {author} {\bibfnamefont {O.}~\bibnamefont
  {Halpern}}\ and\ \bibinfo {author} {\bibfnamefont {T.}~\bibnamefont
  {Holstein}},\ }\href@noop {} {\bibfield  {journal} {\bibinfo  {journal}
  {Phys. Rev.}\ }\textbf {\bibinfo {volume} {59}},\ \bibinfo {pages} {960}
  (\bibinfo {year} {1941})}\BibitemShut {NoStop}%
\bibitem [{\citenamefont {Mitsuda}\ \emph {et~al.}(1992)\citenamefont
  {Mitsuda}, \citenamefont {Yoshizawa},\ and\ \citenamefont
  {Endoh}}]{mitsuda1992neutron}%
  \BibitemOpen
  \bibfield  {author} {\bibinfo {author} {\bibfnamefont {S.}~\bibnamefont
  {Mitsuda}}, \bibinfo {author} {\bibfnamefont {H.}~\bibnamefont {Yoshizawa}},\
  and\ \bibinfo {author} {\bibfnamefont {Y.}~\bibnamefont {Endoh}},\
  }\href@noop {} {\bibfield  {journal} {\bibinfo  {journal} {Phys. Rev. B}\
  }\textbf {\bibinfo {volume} {45}},\ \bibinfo {pages} {9788} (\bibinfo {year}
  {1992})}\BibitemShut {NoStop}%
\bibitem [{\citenamefont {Mydosh}(1993)}]{mydosh1993spin}%
  \BibitemOpen
  \bibfield  {author} {\bibinfo {author} {\bibfnamefont {J.~A.}\ \bibnamefont
  {Mydosh}},\ }\href@noop {} {\emph {\bibinfo {title} {Spin glasses: an
  experimental introduction}}}\ (\bibinfo  {publisher} {CRC Press},\ \bibinfo
  {year} {1993})\BibitemShut {NoStop}%
\bibitem [{\citenamefont {Mondal}\ \emph {et~al.}(2020)\citenamefont {Mondal},
  \citenamefont {Dan}, \citenamefont {Mondal}, \citenamefont {Bhowmik},
  \citenamefont {Ranganathan},\ and\ \citenamefont
  {Mazumdar}}]{mondal2020identification}%
  \BibitemOpen
  \bibfield  {author} {\bibinfo {author} {\bibfnamefont {B.}~\bibnamefont
  {Mondal}}, \bibinfo {author} {\bibfnamefont {S.}~\bibnamefont {Dan}},
  \bibinfo {author} {\bibfnamefont {S.}~\bibnamefont {Mondal}}, \bibinfo
  {author} {\bibfnamefont {R.}~\bibnamefont {Bhowmik}}, \bibinfo {author}
  {\bibfnamefont {R.}~\bibnamefont {Ranganathan}},\ and\ \bibinfo {author}
  {\bibfnamefont {C.}~\bibnamefont {Mazumdar}},\ }\href@noop {} {\bibfield
  {journal} {\bibinfo  {journal} {Intermetallics}\ }\textbf {\bibinfo {volume}
  {120}},\ \bibinfo {pages} {106740} (\bibinfo {year} {2020})}\BibitemShut
  {NoStop}%
\bibitem [{\citenamefont {Pakhira}\ \emph {et~al.}(2016)\citenamefont
  {Pakhira}, \citenamefont {Mazumdar}, \citenamefont {Ranganathan},
  \citenamefont {Giri},\ and\ \citenamefont {Avdeev}}]{pakhira2016large}%
  \BibitemOpen
  \bibfield  {author} {\bibinfo {author} {\bibfnamefont {S.}~\bibnamefont
  {Pakhira}}, \bibinfo {author} {\bibfnamefont {C.}~\bibnamefont {Mazumdar}},
  \bibinfo {author} {\bibfnamefont {R.}~\bibnamefont {Ranganathan}}, \bibinfo
  {author} {\bibfnamefont {S.}~\bibnamefont {Giri}},\ and\ \bibinfo {author}
  {\bibfnamefont {M.}~\bibnamefont {Avdeev}},\ }\href@noop {} {\bibfield
  {journal} {\bibinfo  {journal} {Phys. Rev. B}\ }\textbf {\bibinfo {volume}
  {94}},\ \bibinfo {pages} {104414} (\bibinfo {year} {2016})}\BibitemShut
  {NoStop}%
\bibitem [{\citenamefont {Pakhira}\ \emph {et~al.}(2018)\citenamefont
  {Pakhira}, \citenamefont {Mazumdar}, \citenamefont {Ranganathan},\ and\
  \citenamefont {Giri}}]{pakhira2018magnetic}%
  \BibitemOpen
  \bibfield  {author} {\bibinfo {author} {\bibfnamefont {S.}~\bibnamefont
  {Pakhira}}, \bibinfo {author} {\bibfnamefont {C.}~\bibnamefont {Mazumdar}},
  \bibinfo {author} {\bibfnamefont {R.}~\bibnamefont {Ranganathan}},\ and\
  \bibinfo {author} {\bibfnamefont {S.}~\bibnamefont {Giri}},\ }\href@noop {}
  {\bibfield  {journal} {\bibinfo  {journal} {Journal of Alloys and Compounds}\
  }\textbf {\bibinfo {volume} {742}},\ \bibinfo {pages} {391} (\bibinfo {year}
  {2018})}\BibitemShut {NoStop}%
\bibitem [{\citenamefont {Chakraborty}\ \emph {et~al.}(2022)\citenamefont
  {Chakraborty}, \citenamefont {Gupta}, \citenamefont {Pakhira}, \citenamefont
  {Choudhary}, \citenamefont {Biswas}, \citenamefont {Mudryk}, \citenamefont
  {Pecharsky}, \citenamefont {Johnson},\ and\ \citenamefont
  {Mazumdar}}]{chakraborty2022ground}%
  \BibitemOpen
  \bibfield  {author} {\bibinfo {author} {\bibfnamefont {S.}~\bibnamefont
  {Chakraborty}}, \bibinfo {author} {\bibfnamefont {S.}~\bibnamefont {Gupta}},
  \bibinfo {author} {\bibfnamefont {S.}~\bibnamefont {Pakhira}}, \bibinfo
  {author} {\bibfnamefont {R.}~\bibnamefont {Choudhary}}, \bibinfo {author}
  {\bibfnamefont {A.}~\bibnamefont {Biswas}}, \bibinfo {author} {\bibfnamefont
  {Y.}~\bibnamefont {Mudryk}}, \bibinfo {author} {\bibfnamefont {V.~K.}\
  \bibnamefont {Pecharsky}}, \bibinfo {author} {\bibfnamefont {D.~D.}\
  \bibnamefont {Johnson}},\ and\ \bibinfo {author} {\bibfnamefont
  {C.}~\bibnamefont {Mazumdar}},\ }\href@noop {} {\bibfield  {journal}
  {\bibinfo  {journal} {Physical Review B}\ }\textbf {\bibinfo {volume}
  {106}},\ \bibinfo {pages} {224427} (\bibinfo {year} {2022})}\BibitemShut
  {NoStop}%
\bibitem [{\citenamefont {Mori}\ and\ \citenamefont
  {Mamiya}(2003)}]{mori2003dynamical}%
  \BibitemOpen
  \bibfield  {author} {\bibinfo {author} {\bibfnamefont {T.}~\bibnamefont
  {Mori}}\ and\ \bibinfo {author} {\bibfnamefont {H.}~\bibnamefont {Mamiya}},\
  }\href@noop {} {\bibfield  {journal} {\bibinfo  {journal} {Phys. Rev. B}\
  }\textbf {\bibinfo {volume} {68}},\ \bibinfo {pages} {214422} (\bibinfo
  {year} {2003})}\BibitemShut {NoStop}%
\bibitem [{\citenamefont {Lago}\ \emph {et~al.}(2012)\citenamefont {Lago},
  \citenamefont {Blundell}, \citenamefont {Eguia}, \citenamefont {Jansen},\
  and\ \citenamefont {Rojo}}]{lago2012three}%
  \BibitemOpen
  \bibfield  {author} {\bibinfo {author} {\bibfnamefont {J.}~\bibnamefont
  {Lago}}, \bibinfo {author} {\bibfnamefont {S.}~\bibnamefont {Blundell}},
  \bibinfo {author} {\bibfnamefont {A.}~\bibnamefont {Eguia}}, \bibinfo
  {author} {\bibfnamefont {M.}~\bibnamefont {Jansen}},\ and\ \bibinfo {author}
  {\bibfnamefont {T.}~\bibnamefont {Rojo}},\ }\href@noop {} {\bibfield
  {journal} {\bibinfo  {journal} {Phys. Rev. B}\ }\textbf {\bibinfo {volume}
  {86}},\ \bibinfo {pages} {064412} (\bibinfo {year} {2012})}\BibitemShut
  {NoStop}%
\bibitem [{\citenamefont {Souletie}\ and\ \citenamefont
  {Tholence}(1985)}]{souletie1985critical}%
  \BibitemOpen
  \bibfield  {author} {\bibinfo {author} {\bibfnamefont {J.}~\bibnamefont
  {Souletie}}\ and\ \bibinfo {author} {\bibfnamefont {J.}~\bibnamefont
  {Tholence}},\ }\href@noop {} {\bibfield  {journal} {\bibinfo  {journal}
  {Phys. Rev. B}\ }\textbf {\bibinfo {volume} {32}},\ \bibinfo {pages} {516}
  (\bibinfo {year} {1985})}\BibitemShut {NoStop}%
\end{thebibliography}

\end{document}